\documentclass[10pt,aps,prb,twocolumn,preprintnumbers,amsmath,amssymb,floatfix,showpacs,citeautoscript,superscriptaddress]{revtex4-2}
\usepackage{amsmath}
\usepackage{graphicx}
\usepackage{physics}
\usepackage{bbold}
\usepackage{xcolor}
\usepackage[normalem]{ulem}
\usepackage{soul}
\setstcolor{red} 
\usepackage[colorlinks=true, allcolors=blue]{hyperref}

\begin{document}

\title{Electronic theory for scanning tunneling microscopy spectra in bilayer nickelate thin films}

\author{Marius Scholten}
\affiliation{Theoretische Physik III, Fakult\"at für Physik und Astronomie, Ruhr-Universit\"at Bochum,
  D-44780 Bochum, Germany}

\author{Steffen B\"otzel}
\affiliation{Theoretische Physik III, Fakult\"at für Physik und Astronomie, Ruhr-Universit\"at Bochum,
  D-44780 Bochum, Germany}

\author{Frank Lechermann}
\affiliation{Theoretische Physik III,  Fakult\"at für Physik und Astronomie, Ruhr-Universit\"at Bochum,
  D-44780 Bochum, Germany}

\author{Peayush Choubey}
\affiliation{Department of Physics, Indian Institute of Technology Roorkee, Roorkee 247667, Uttarakhand, India}

\author{Ilya M. Eremin}
\affiliation{Theoretische Physik III,  Fakult\"at für Physik und Astronomie, Ruhr-Universit\"at Bochum,
  D-44780 Bochum, Germany}

\begin{abstract}
Recent Scanning Tunneling Microscopy (STM) experiments measuring the superconducting gap features in thin films of superconducting bilayer nickelates La$_2$PrNi$_2$O$_7$ at ambient pressure and compressive strain paved the way to study the Cooper-pairing models and the band-selective identification of the gap features in these systems. Here, using the realistic two-orbital bilayer model and the continuum Green's function formalism, we theoretically analyze orbital and band-selective local density of states as well as the corresponding STM spectra. We find that the multiorbital character and the spatial dependence of the Wannier functions leads to the spectra developing characteristic features depending on the position of the scanning tunneling microscope's tip. This allows for a band-resolved analysis of the superconducting coherence peaks and scattering momenta. We identify a clear path for experimental measurements to not only identify the debated incipiency of the $\gamma$-band, but also identification of the coherence peaks' band origins via distance dependent measurements of the local density of states and its corrections through impurity scattering.
\end{abstract}

\maketitle
\section{Introduction}
\label{sec:Intro}
\begin{figure}[t]
	\includegraphics[width=0.48\linewidth]{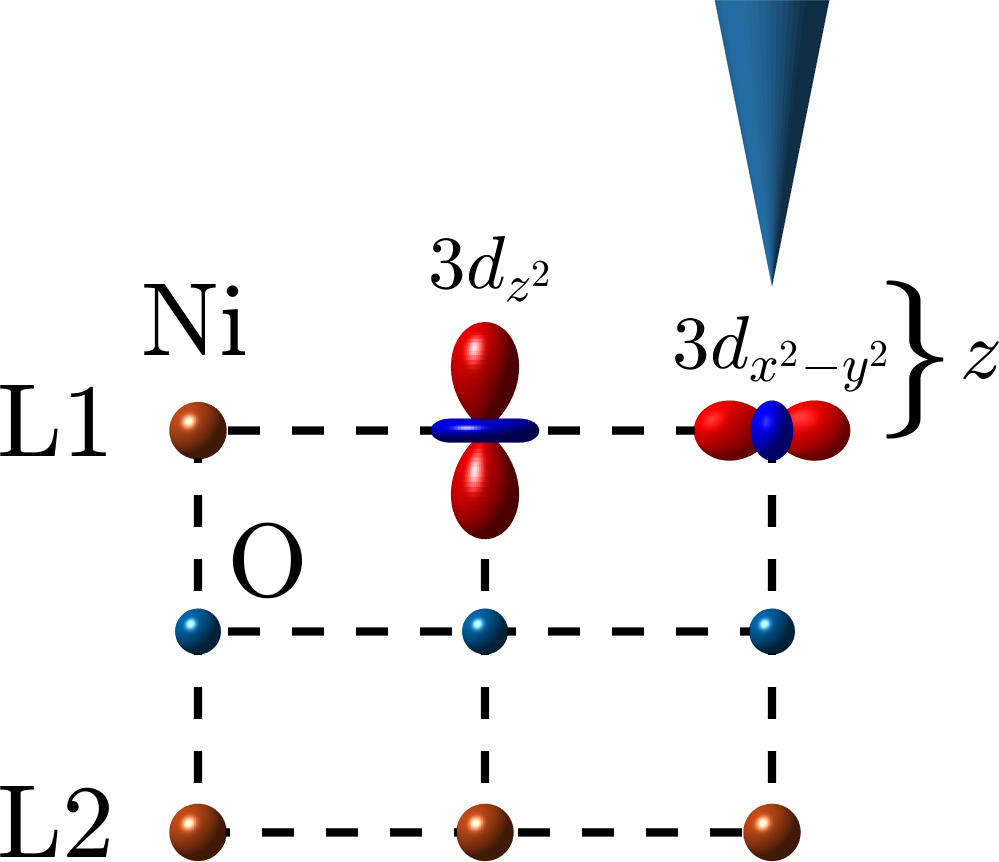}\put(-120,72){\textrm{(a)}}\hspace{1pt}
    \includegraphics[width=0.48\linewidth]{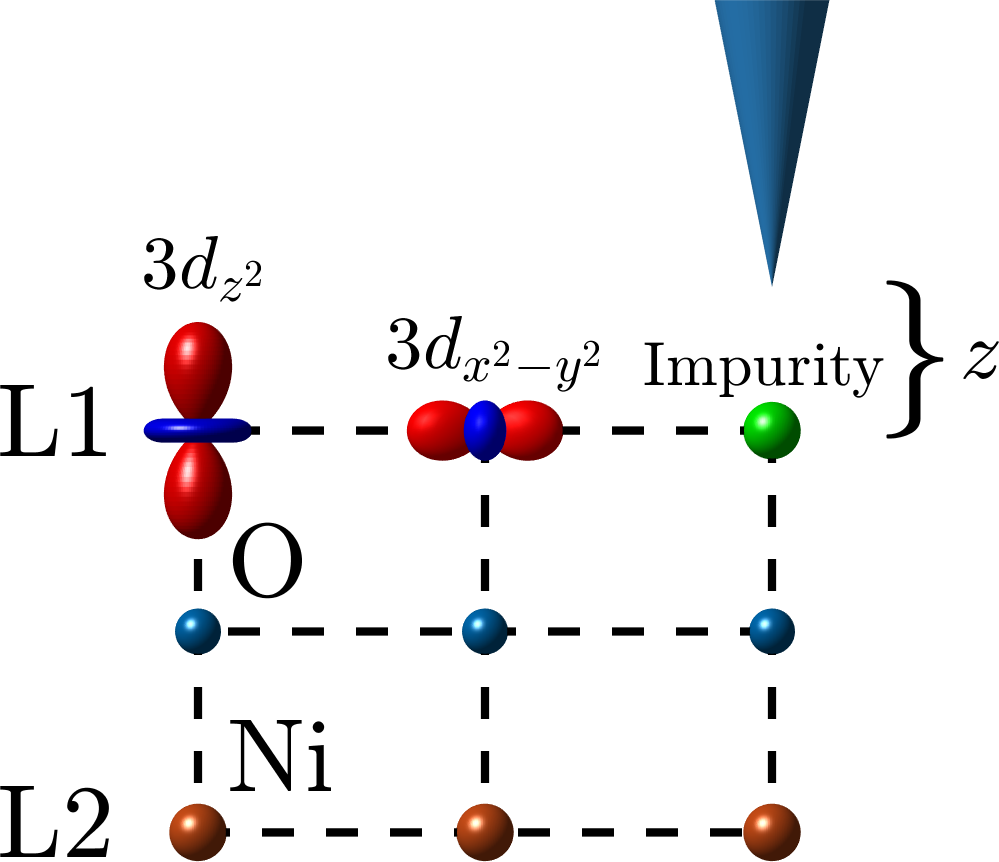}\put(-120,72){\textrm{(b)}}\hspace{1pt}
	\caption{Illustration for our two-orbital bilayer nickelate model. The Ni atoms host the $3d_{z^2}$ and $3d_{x^2-y^2}$ orbitals. The STM tip is placed above the upper layer L1. Depending on the configuration of the experiment the current through the tip can either tunnel into the clean surface without any impurities  (a) or via the impurity on the top layer if the tip is placed above the impurity site (b). The impurity can also be located on the apical oxygen site or in the bottom layer L2.}
\label{QPIConceptFig}
\end{figure}

The recent discovery of high-temperature superconductivity in pressurized La$_3$Ni$_2$O$_7$ (La-327)~\cite{sun23,JunHou:117302,zhang2023high,zhou2023evidence,zhang2023effects,wang2023structure,wang2023pressure,dong2023visualization} and La$_2$PrNi$_2$O$_7$~\cite{WangNature24}, as well as in thin films of La$_{3-x}$Pr${_x}$Ni$_2$O$_7$ $(x = 0,0.15,1)$ grown on compressively strained SrLaAlO$_4$(001) substrates~\cite{ko2024signatures,zhou2025ambient,bhatt2025resolvingstructuraloriginssuperconductivity,liu2025superconductivitynormalstatetransportcompressively}, has revived strong interest in bilayer superconducting systems. In these materials, multiorbital physics and interlayer coupling lead to a complex low-energy electronic structure, setting them apart from cuprates and giving rise to multiple competing pairing channels.

On the theoretical side, both interlayer-driven $s_{\pm}$-wave pairing and intralayer spin-fluctuation–driven $d$-wave states have been proposed within multiorbital bilayer models~\cite{lechermann2023electronic,liu2023role,qin2023high,huang2023impurity,oh2023type,qu2023bilayer,ryee2024quenched,tian2023correlation,liao2023electron,kaneko2023pair,LuoModel,chen2023orbital,jiang2023high,shen2023effective,yang2023minimal,wu2023charge,luo2024high,Zhan-Hu-2025-Bilayer-FRG-ElectronPhonon,Le-Hu-2025-StrainedBilayerThin-SDW,Zhan-Hu-2025-Bilayer-FRG-NonlocalCoulomb-SDW-CDW,Maier-Dagotto-2025-Bilayer-InterlayerPairing-Sus,Ushio-Kuroki-2025-FLEX-LinGapSus-Spm,Dao-Xin-2025-S+-Pairing-RMFT,ryee2025superconductivity}. This competition reflects the interplay between intra- and interlayer pairing tendencies, similar to earlier bilayer Hubbard model studies~\cite{bulut1992nodeless,Maier2011,Vanhala2015,Maier2019,Matsumoto2020,Karakuzu2021}. However, a consistent experimental determination of the superconducting gap symmetry and the bands involved in bilayer nickelates remains lacking.
Existing experiments provide partial and sometimes conflicting information. Point-contact spectroscopy on bulk samples~\cite{cao2025direct,liu2025andreev,guo2025revealing} and angle-resolved photoemission spectroscopy (ARPES) on thin films~\cite{Shen-Chen-2025-ARPES-S+-,li2025photoemissionevidencemultiorbitalholedoping,wang2025electronic,li2026threedimensionalelectronicstructuressuperconducting} support different superconducting gap symmetry candidates. At the same time recent scanning tunneling microscopy (STM) measurements reveal two superconducting gaps associated with different bands and are compatible with a multiband $s$-wave scenario~\cite{Hai-Hu-Wen-2025-STMSpm,wang-2026-STM-LaPr-Thin-UShape2Peaks,liang-2026-STM-LaPr-Thin-UShape}. These results highlight the urgent need to disentangle gap symmetry and band selectivity in a unified framework.

In this work, we address this issue by performing realistic simulations of STM observables employing first-principles Wannier functions to properly account for the tunneling processes within the continuum Green’s function approach \cite{choubey2014visualization,kreisel2015interpretation,Choubey2017B,Kreisel2016LiFeAs}, see Fig.~\ref{QPIConceptFig}. This approach yields local density of states (LDOS) at the STM tip position (usually, a few angstroms above the sample) which, in turn, is proportional to the differential tunneling conductance measured by the STM apparatus \cite{TersoffHamann1985}. Unlike lattice-based calculations, which yield LDOS at lattice sites only, the continuum Greens function method correctly captures the local symmetry at the STM tip position. In multiorbital systems such as bilayer nickelates, the continuum Green’s function approach becomes particularly powerful \cite{Choubey2021Nickelate} because the wave function of the STM tip couples differently to the orbital space matrix, consisting of 3$d_{x^2-y^2}$- and 3$d_{z^2}$-orbitals, which also hybridize with planar oxygen $p_x$, $p_y$ orbitals and apical $p_z$ orbitals, respectively. This orbital selectivity allows to disentangle various orbital contributions. We first compare the candidate $s_{\pm}$- and $d$-wave pairing states within an effective bilayer model and analyze possible gap distribution on the involved orbitals assuming the tunneling current into the clean surface, as shown in Fig.~\ref{QPIConceptFig}(a). Our goal is to determine which gap symmetries and band assignments are consistent with existing experiments and to identify robust signatures in the LDOS.
Here, we demonstrate that tip height-dependent measurements of tunneling spectra can clearly distinguish between the coherence peaks' band origin and the actual symmetry of the supercoducting order parameter. Secondly, we study the response of an isolated impurity, substituting a Ni atom in one of the NiO layers or an apical oxygen atom between the two layers (see  Fig.~\ref{QPIConceptFig}(b)), via the $T$-matrix approach. Here, we show that the quasiparticle interference (QPI) patterns measured in STM experiments can unambiguously distinguish between the candidate superconducting structures.

The manuscript is organized as follows. In Section \ref{sec:Model}, we introduce a two-orbital, bilayer tight-binding model for La-327 derived from first-principles calculations and account for superconductivity via a mean-field pairing Hamiltonian, considering $s_{\pm}$- and $d$-wave superconducting gap structures. Next, in Section \ref{sec:LDOS}, we predict distinct STM tunneling spectra for each of the candidate gap structures by calculating the continuum local density of states spectra in a clean system. In Section \ref{sec:QPI}, we calculate impurity-induced corrections to the LDOS, highlighting distinct QPI signatures of $s_{\pm}$- and $d$-wave gap structures. Finally, we summarize our findings in Section \ref{sec:Summary}.

\section{Model}
\label{sec:Model}
\begin{figure}
	\includegraphics[width=0.49\linewidth]{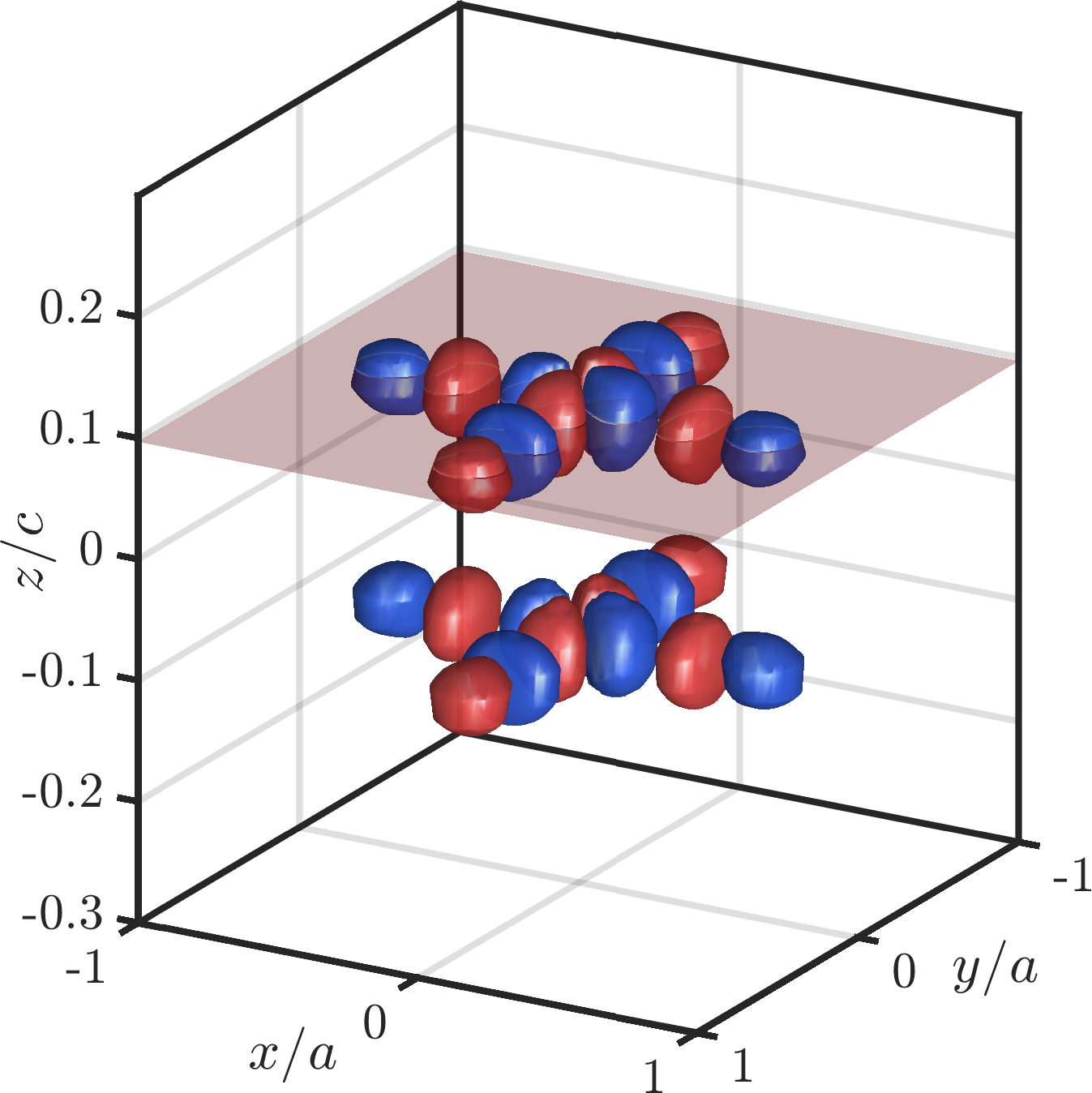}\put(-120,112){\textrm{(a)}}\hspace{1pt}
    \includegraphics[width=0.49\linewidth]{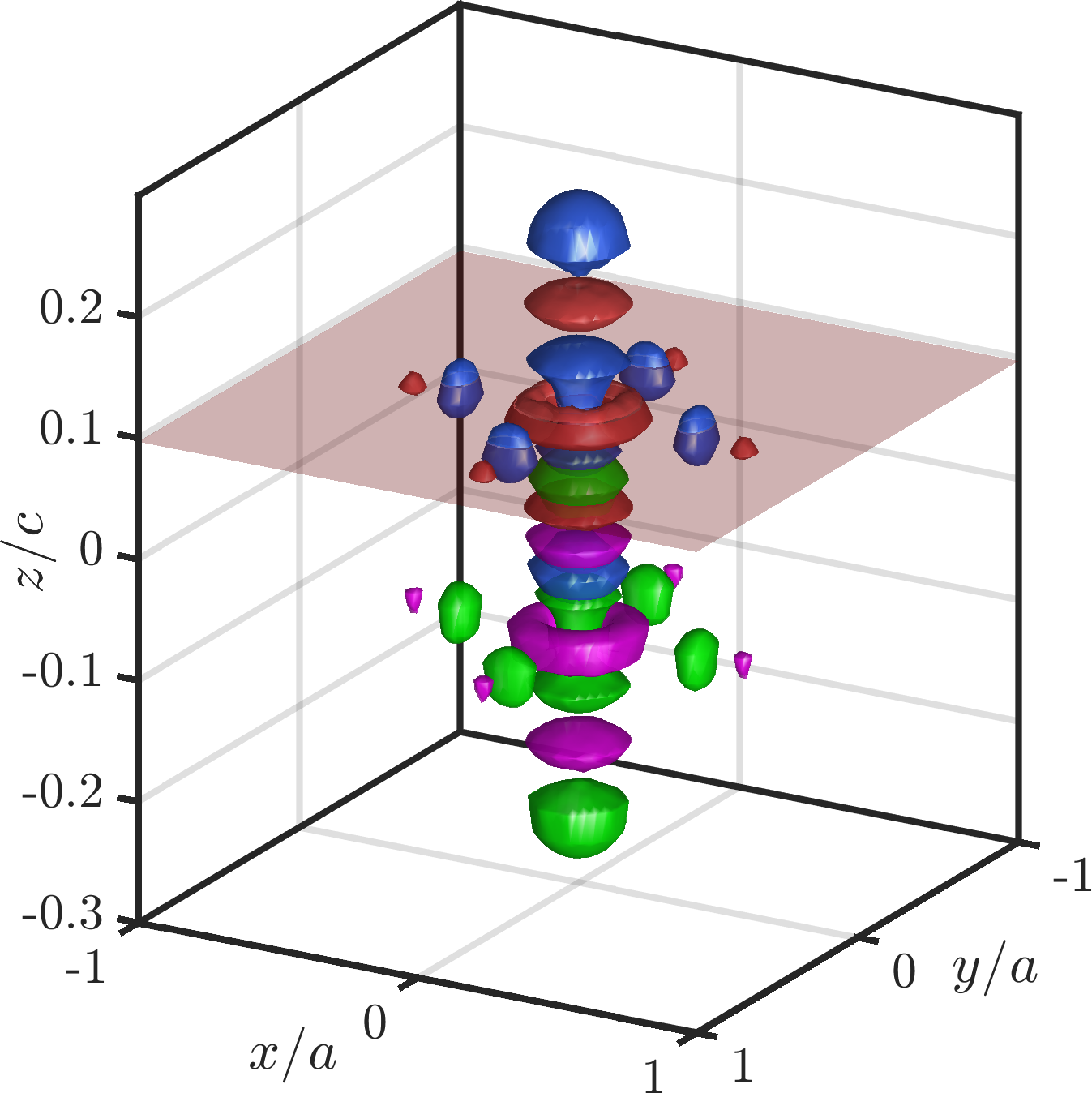}\put(-110,112){\textrm{(b)}}\\
    \includegraphics[width=0.49\linewidth]{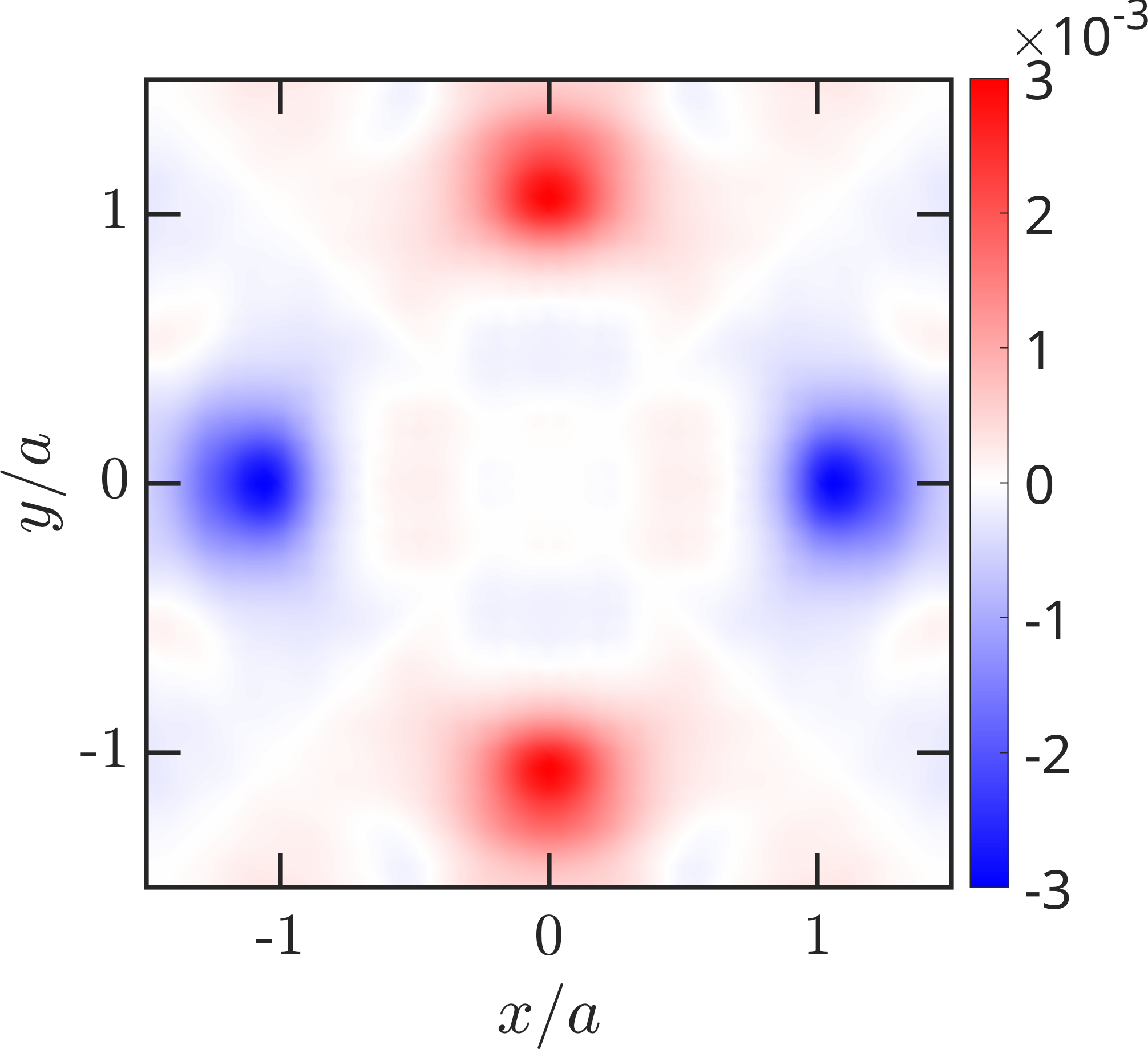}\put(-120,112){\textrm{(c)}}\hspace{1pt}
    \includegraphics[width=0.49\linewidth]{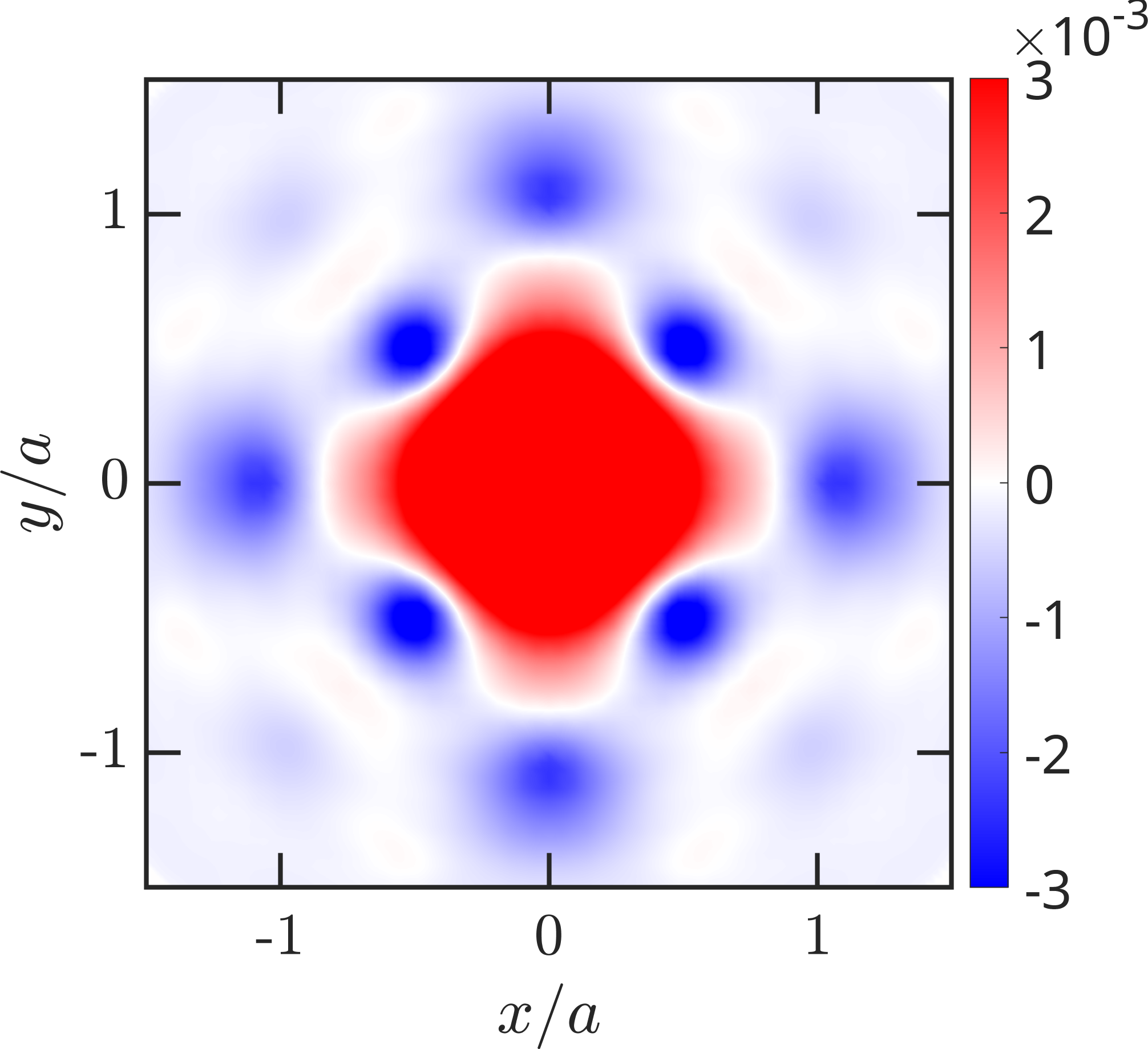}\put(-110,112){\textrm{(d)}}
	\caption{Three dimensional isosurface plots of the $3d_{x^2-y^2}$(a) and $3d_{z^2}$(b) Wannier orbitals employed in this work, shown at 5\% of their maximum value for both layers. In (b) the coloring of the Wannier orbital on the bottom layer differs to clarify the individual contributions. (c) and (d) show the $z$-cut of the $3d_{x^2-y^2}$ and $3d_{z^2}$ Wannier orbitals at a distance of $3\,\textrm{\AA}$ away from the top layer indicated by the red plane in (a) and (b).}
\label{Wannier3dAndCutsFig}
\end{figure}

\begin{figure}
	\includegraphics[width=0.9\linewidth]{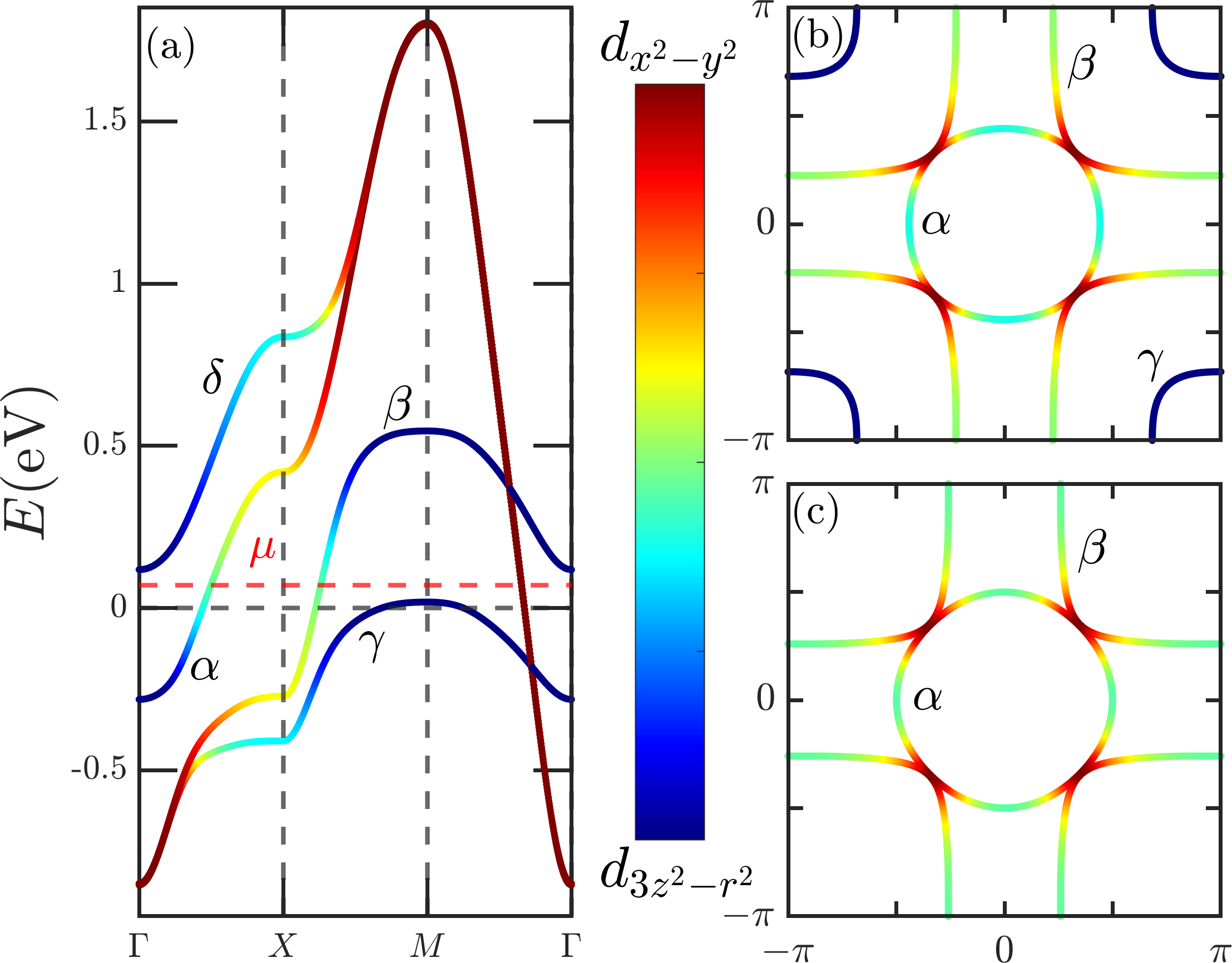}
	\caption{Calculated electronic dispersion (a) along the high-symmetry path and the resulting Fermi surface (b,c) colored according to the orbital weight for the low-energy model~\cite{Lechermann-2025-ThinFilmSlaveBoson,Lechermann-2007-SlaveBosonFormalism}, employed in our study. Following the debate in the literature~\cite{li2025photoemissionevidencemultiorbitalholedoping,wang2025electronic,Shen-Chen-2025-ARPES-S+-}, we consider the cases of $\gamma$-band either crossing (b) or located $\approx$ 50\,meV (c) below the Fermi level.}
\label{DispNormal}
\end{figure}

The starting point to model our bilayer system is the non-interacting Hamiltonian of the form
\begin{equation}
    \begin{split}
        \mathcal{H}_0=\sum_{\bf{k}\ell\ell'\mu\nu\sigma} H_{0,\ell\mu;\ell'\nu}({\bf{k}}) c_{\bf{k}\sigma,\ell\mu}^\dagger c_{\bf{k}\sigma,\ell'\nu}^{\phantom{\dagger}},
    \end{split}
\end{equation}
where $c_{\bf{k}\sigma,\ell\mu}^\dagger$ creates an electron with momentum ${\bf{k}}=(k_x,k_y)$, spin $\sigma$, in layer $\ell$ and Ni-$3d$ orbital $\mu$. The matrix elements of the tight-binding Hamiltonian $H_{0,\ell\mu;\ell'\nu}(\bf{k})$ are described by the maximally-
localized Wannier orbital projection of the DFT band structure~\cite{lechermann2023electronic,LuoModel,li2025theoreticalstudyelectronicproperties} on the $3d_{x^2-y^2}$- and $3d_{z^2}$-orbitals in each layer, including the slave-boson renormalization of the hopping integrals, performed for thin film model\cite{Lechermann-2025-ThinFilmSlaveBoson,Lechermann-2007-SlaveBosonFormalism}. Note that the hybridization with neighboring oxygen $p_x$, $p_y$ and $p_z$ orbitals is effectively included by the extended character of the Wannier orbitals as outlined in Fig.~\ref{Wannier3dAndCutsFig}.
%
The two-orbital bilayer Hamiltonian is also mirror symmetric with respect to layer indices yielding its block structure
\begin{equation}
    \begin{split}
        \hat{H}_{0}({\bf{k}})=\mqty(\hat{H}^{\parallel}_{\bf{k}}&\hat{H}^{\perp}_{\bf{k}}\\\hat{H}^{\perp}_{\bf{k}}&\hat{H}^{\parallel}_{\bf{k}}),
    \end{split}
    \label{HamiltonianNonInt}
\end{equation}
with intralayer and interlayer subblocks $\hat{H}^{\parallel}_{\bf{k}}$ and $\hat{H}^{\perp}_{\bf{k}}$, respectively. We introduce the bonding–antibonding transformation $c^{\textrm{b/a}}_{{\bf{k}}\sigma,\mu}=\frac{1}{\sqrt{2}}(c_{\bf{k}\sigma,1\mu}\pm c_{{\bf{k}}\sigma,2\mu})$, which block-diagonalizes the Hamiltonian and yields bonding and antibonding bands. The resulting quasiparticle dispersion in the normal state, consisting of four bands, is shown in Fig.~\ref{DispNormal}, where the orbital weight is indicated by the color scale. The model hosts three bands crossing the Fermi level, labeled $\alpha$, $\beta$, and $\gamma$. The $\alpha$- and $\gamma$-bands are bonding, while the $\beta$- and $\delta$-bands are antibonding; the latter does not cross the Fermi level. Note, the precise position of the $\gamma$-band relative to the Fermi level remains under debate~\cite{li2025photoemissionevidencemultiorbitalholedoping,wang2025electronic,Shen-Chen-2025-ARPES-S+-}. To account for this uncertainty, we also consider the case of an incipient $\gamma$-band by introducing a slight chemical potential shift, as shown in Fig.~\ref{DispNormal}(a).
The superconducting state can be included straightforwardly on a mean-field level by expanding the bilayer basis to the Bogoliubov-Nambu space, leading to the following Cooper-pairing Hamiltonian
\begin{equation}
    \begin{split}
   \mathcal{H}_{\textrm{pair}}=\sum_{\bf{k}}\mathcal{D}_{\ell\mu;\ell'\nu}({\bf{k}}) c_{\bf{k}\uparrow,\ell\mu}^\dagger c_{-\bf{k}\downarrow,\ell'\nu}^{\dagger}+\textrm{h.c.}
    \label{PairingHam}
    \end{split}
\end{equation}
\begin{equation}
    \begin{split}
   \hat{\mathcal{D}}({\bf{k}})=\mqty(\hat{\Delta}^{\parallel}_{\bf{k}}&\hat{\Delta}^{\perp}_{\bf{k}}\\\hat{\Delta}^{\perp}_{\bf{k}}&\hat{\Delta}^{\parallel}_{\bf{k}}).
    \label{PairingHam2}
    \end{split}
\end{equation}
As has been discussed previously by various groups, ~\cite{lechermann2023electronic,liu2023role,qin2023high,huang2023impurity,oh2023type,qu2023bilayer,ryee2024quenched,tian2023correlation,liao2023electron,kaneko2023pair,LuoModel,chen2023orbital,jiang2023high,shen2023effective,yang2023minimal,wu2023charge,luo2024high,Zhan-Hu-2025-Bilayer-FRG-ElectronPhonon,Le-Hu-2025-StrainedBilayerThin-SDW,Zhan-Hu-2025-Bilayer-FRG-NonlocalCoulomb-SDW-CDW,Maier-Dagotto-2025-Bilayer-InterlayerPairing-Sus,Ushio-Kuroki-2025-FLEX-LinGapSus-Spm,Dao-Xin-2025-S+-Pairing-RMFT}
the main superconducting instabilities vary between $d_{x^2-y^2}$- (or $d_{xy}$) wave and bonding-antibonding $s_\pm$-wave symmetries where the latter seems to dominate in most scenarios. As was argued previously \cite{Boetzel2024} the result may depend sensitively on the relative strength of the intra- versus interlayer antiferromagnetic spin fluctuations. Here, we consider both scenarios, but vary the gap magnitudes on each orbital (band) so that it fits the position of the coherence peaks, as found in recent STM measurements \cite{Hai-Hu-Wen-2025-STMSpm}. In particular, the observed spectrum exhibits two superconducting gap magnitudes of approximately $7\,$meV and $19\,$meV. 

In what follows, we assume the Cooper-pairing to take a simple intraorbital form, where we define the individual components outlined by Eq. (\ref{PairingHam2}) as $(\hat{\Delta}^{\parallel/\perp}_{\bf{k}})_{\mu\nu}=\delta_{\mu\nu}\Delta^{\parallel/\perp}_{\mu,0}\gamma_{\bf{k}}$, where $\gamma_{\bf{k}}$ is the symmetry factor of the underlying gap symmetry, i.e. $\gamma_{\bf{k}}=1$ for the $s$-wave and $\gamma_{\bf{k}}=(\cos(k_x)-\cos(k_y))/2$ for the $d$-wave. 
Note that the distribution of gap magnitude on the 
$\alpha$, $\beta$, and $\gamma$-bands, found in ARPES \cite{Shen-Chen-2025-ARPES-S+-}, would be indicative of the dominant superconducting gap to be on the $d_{z^2}$-orbital, as the $\gamma$-Fermi surface sheet has a dominant $d_{z^2}$-orbital contribution and the $\beta$-Fermi surface sheet pairing takes its minimum value along the diagonals, where the $d_{x^2-y^2}$-orbital is the dominant, see Fig.~\ref{DispNormal}.  The largest superconducting gap on the $d_{z^2}$-orbital has also been found theoretically by multiple groups within the $s_{\pm}$-wave scenario  \cite{liu2023role,lechermann2023electronic,yang2023possible,luo2024high,oh2023type,huang2023impurity,yang2023strong,sakakibara2024possible,Hai-Hu-Wen-2025-STMSpm,Dao-Xin-2025-S+-Pairing-RMFT,Shen-Chen-2025-ARPES-S+-}.
Nevertheless, there has been a proposal \cite{Khaliullin-2026-triplonSCgap} where the largest superconducting gap was found on the $\alpha$-Fermi surface pocket, and we model the STM spectra for this situation as well. Appendix \ref{appendix_SuperconductingGaps} lists values of $\Delta^{\parallel/\perp}_{\mu,0}$ for all the situations considered in this work. 

\section{Local density of states and STM spectra}
\label{sec:LDOS}
Using the resulting Bogoliubov-de Gennes (BdG) Hamiltonian in Nambu space  $\mathcal{H}=\mathcal{H}_{0}+\mathcal{H}_{\textrm{pair}}$ we define the bare Green's function as $\hat{G}_{\bf{k}}(\omega)=[(\omega+i\delta)-\hat{H}_{\bf{k}}]^{-1}$, where $\hat{H}_{\bf{k}}$ is the resulting BdG matrix.
The lattice density of states, projected onto the outer layer of our system could be calculated via the standard expression
\begin{equation}
	\begin{split}
	   \rho(\omega)=-\frac{2}{\pi}\sum_{\bf{k}} \mathrm{Im}  \left[\mathrm{Tr} \frac{\tau_0+\tau_z}{2}\hat{L}_1\hat{G}_{\bf{k}}(\omega) \right],
	\end{split}
	\label{DOSLatticeFormula}
\end{equation}
where the factor 2 results from tracing over spin space, $\tau_i$ are Pauli matrices in particle-hole space and $\hat{L}_1$ is the projection operator onto the upper layer. 

However, to enable a meaningful comparison between theoretically calculated quantities and experimental STM measurements, it is essential to accurately determine the position of the STM tip relative to the underlying bilayer lattice structure. In contrast to the layer projection $\hat{L}_1$ of the lattice calculation, the tip position, which is located outside of the sample, can be simulated more precisely by computing the continuum Green's function
\begin{equation}
	\begin{split}
	   \hat{\mathcal{G}}({\bf{r}},{\bf{r}}',\omega)=\sum_{{\bf{R}\bf{R}}'\ell\ell'\mu\nu} w^{\ell\mu}_{\bf{R}}({\bf{r}}) \hat{\mathcal{G}}_{{\bf{R}\bf{R}}'}^{\ell\mu;\ell'\nu}(\omega) w^{\ell'\nu*}_{\bf{R}'}({\bf{r}}')\\
	\end{split}
	\label{ContGFDef}
\end{equation}
where $w^{\ell\mu}_{\bf{R}}({\bf{r}})$ is the Wannier function of layer $\ell$ and orbital $\mu$ with lattice vector $\bf{R}$ and continuum position $\bf{r}$. The Wannier functions employed in this analysis are obtained numerically by the standard transformation of $k$-resolved DFT Bloch functions into associated real-space functions according to the maximally-localized Wannier-construction scheme~\cite{marzari12}. We define the lattice Green's function $\hat{\mathcal{G}}_{{\bf{R}\bf{R}}'}(\omega)=\hat{G}_{{\bf{R}}-{\bf{R}'}}(\omega)+\hat{G}_{{\bf{R}}}(\omega)\hat{T}(\omega)\hat{G}_{{-\bf{R}'}}(\omega)$ in which $G_{{\bf{R}}}(\omega)$ represents the real-space analogue of the momentum-space Green's function $G_{{\bf{k}}}(\omega)$ discussed previously. With the continuum Green's function we can now define the continuum LDOS at the STM tip position $\bf{r}$ as
\begin{equation}
    \rho({\bf{r}},\omega)=-\frac{2}{\pi}\Im{\hat{\mathcal{G}}^{\textrm{11}}({\bf{r}},{\bf{r}},\omega)}
	\label{LDOSContGFFormula}
\end{equation}
where the exponent 11 denotes the usual projection to the particle channel in Nambu space. The projection onto the Wannier orbitals now plays a similar role to our previous projection operator $\hat{L}_1$ in Eq. (\ref{DOSLatticeFormula}), capturing the positioning of the STM tip above the sample and now also including the orbital structure of the bilayer system.
Moreover, since the system's Wannier functions reach over multiple unit cells, they naturally include intra-unit-cell contributions, which provides enhanced momentum-space resolution compared to discrete lattice models, in which the allowed wave vectors are limited by the lattice constant $a$ to the first Brillouin zone $k_i\in[-\frac{\pi}{a},\frac{\pi}{a}]$.

The resulting Wannier functions of the bilayer system are displayed in Fig.~\ref{Wannier3dAndCutsFig}. Panels (a) and (b) present three-dimensional isosurface plots for the Wannier functions of each layer and orbital. Here, (a) shows the $3d_{x^2-y^2}$-orbitals and (b) presents the $3d_{z^2}$-orbitals. The transparent red plane marks the position of the top layer, corresponding to $z=0$, the height around which the STM tip is approximately placed. Panels (c) and (d) show two-dimensional cuts taken at a distance of $z\approx3\,\textrm{\AA}$ from this red plane. From these cuts and the corresponding color scale, it is evident that the $3d_{x^2-y^2}$-orbital shown in panel (c) exhibits smaller maximum and minimum weights compared to the $3d_{z^2}$ orbital shown in (d). The maximum value of the $3d_{z^2}$ orbital is cut off leading to a high intensity feature localized at the center. If one were to plot it on an individual scale, the $3d_{z^2}$ orbital would appear as an almost atomic-like distribution, whereas comparatively the $3d_{x^2-y^2}$ orbital has a larger spatial broadening.
An additional feature visible in the three-dimensional representations is the pronounced extension of the $3d_{z^2}$ orbital along the $z$-direction. As a consequence, the corresponding Wannier functions extend much further away from the top layer, effectively acting as a $3d_{z^2}$-orbital filter at larger distances above the surface. 

In the following we begin by considering the DOS calculated on the lattice level and further advance to the continuum version later on. We inspect the density of states for the different scenarios of an incipient and crossing $\gamma$-band.

\subsection{Incipient $\gamma$-band}
\label{sec:Incipient_Gamma}
\begin{figure}
	\includegraphics[width=0.99\linewidth]{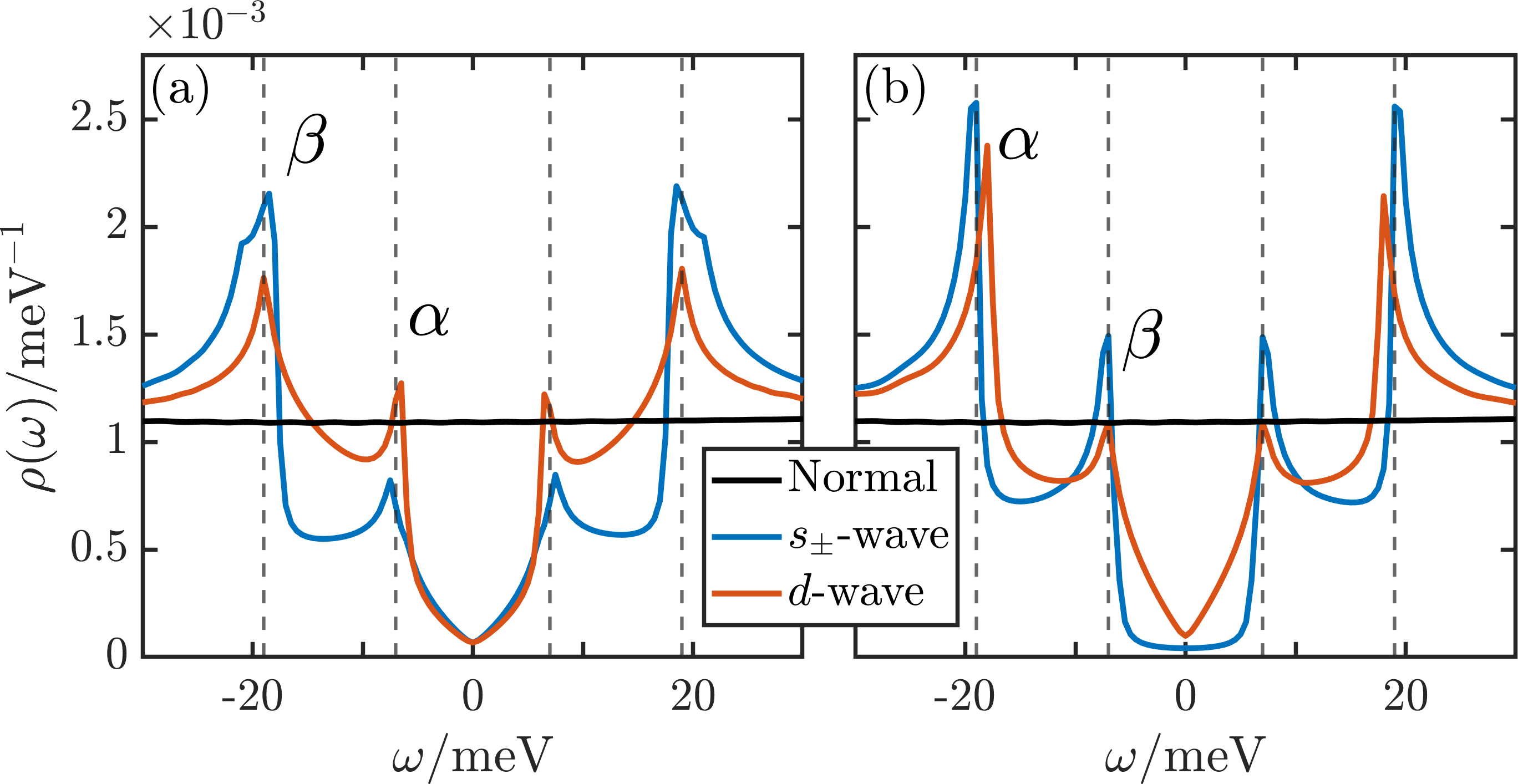}
	\caption{(a) DOS calculated on the lattice in absence of any impurities for the incipient $\gamma$-band case and for the two candidate gap symmetries portrayed in Fig.~\ref{FigGapOnBandsDiffSym}, where the $\beta$-band produces a coherence peak around $19\,$meV and the $\alpha$-band at $7\,$meV. (b) Vice versa situation where the $\alpha$-band peak lies at $19\,$meV and the $\beta$-band peak at $7\,$meV. The dashed lines are guides to the $7$ and $19\,$meV values reported in\cite{Hai-Hu-Wen-2025-STMSpm}.}
\label{DOSLatticeInciFig1}
\end{figure}

In the incipient $\gamma$-band case Fig.~\ref{DOSLatticeInciFig1} (a) shows the lattice DOS calculated for the gap structures shown in Fig.~\ref{FigGapOnBandsDiffSym}, where the coherence peak around 19 meV corresponds to the gap on the $\beta$-band, while the low energy peak is owed to the $\alpha$-band. Fig.~\ref{DOSLatticeInciFig1} (b) shows the inverse scenario, where the $\alpha$-band produces the higher energy peak and the $\beta$-band produces the low energy signal. In this case the projected gap has a similar sign structure as portrayed in Fig.~\ref{FigGapOnBandsDiffSym}, with the only differences being the gap magnitudes swapping and the $\alpha$-band no longer hosting accidental nodes for the $s_\pm$-wave. Obviously, within lattice Green's functions both symmetries and both scenarios for the gap size of the $\alpha$- and $\beta$-band show significant spectral weight for the low energy coherence peak around $7\,$meV, which contrasts the experimental measurement\cite{Hai-Hu-Wen-2025-STMSpm,liang-2026-STM-LaPr-Thin-UShape,wang-2026-STM-LaPr-Thin-UShape2Peaks}. 

\begin{figure}
	\includegraphics[width=1\linewidth]{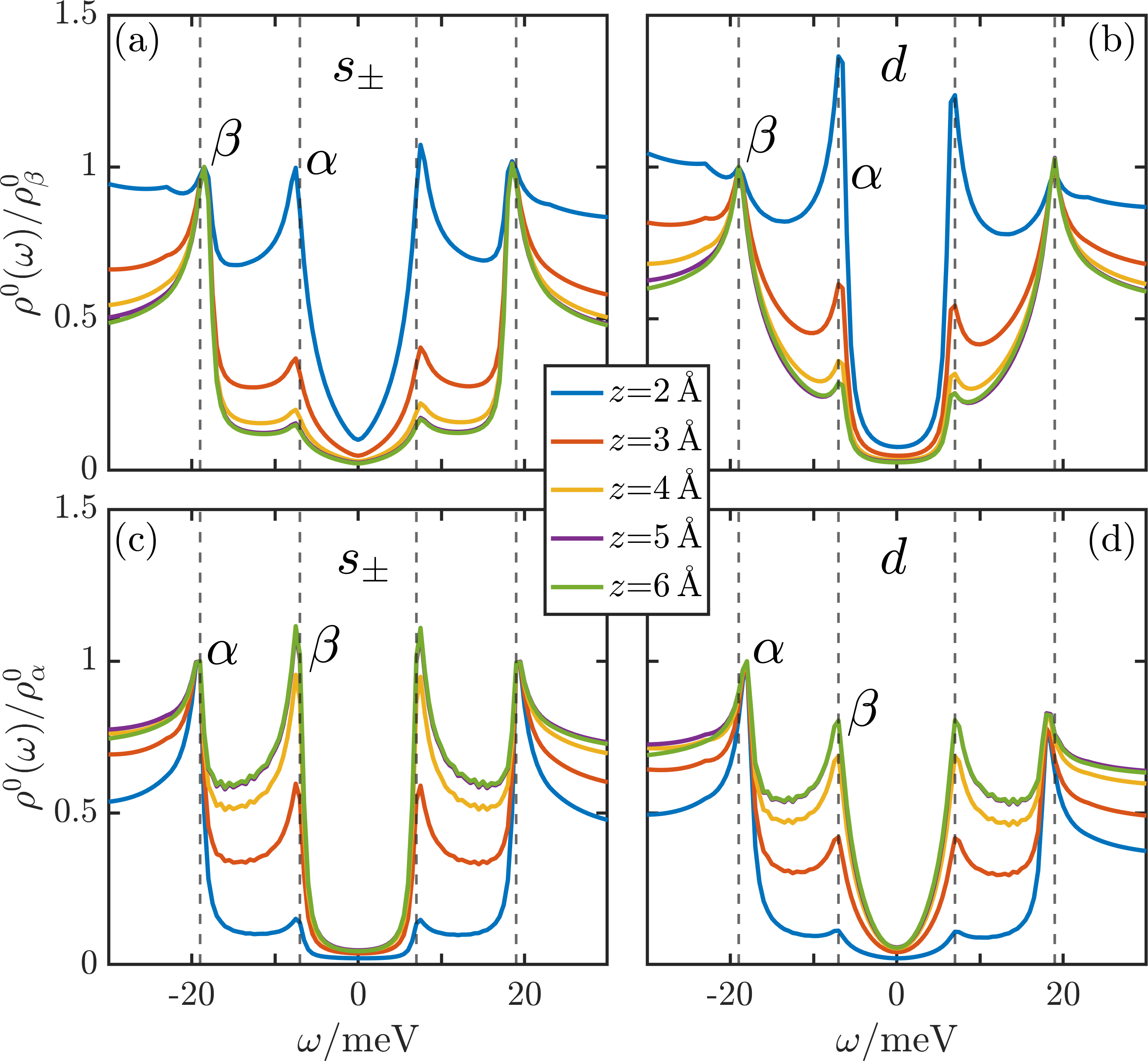}
	\caption{(a)+(b) LDOS normalized by the left shoulder $\beta$-band coherence peak calculated for the incipient $\gamma$-band cases shown in Fig.~\ref{FigGapOnBandsDiffSym} with larger $\beta$-band gap compared to the $\alpha$-band, and without any impurities in dependency on the distance from the top layer of the sample located at $z=0$. The STM tip is directly placed over the lattice site. (c)+(d) Show the reversed situation of a smaller $\beta$-band gap compared to the $\alpha$-band. (a)+(c) Show the evolution of the LDOS for the $s_\pm$-wave symmetry while (b)+(d) show the $d$-wave symmetry.}
\label{FigDOS0_sandd_Wannier}
\end{figure}

Switching to the Wannier function description and continuum Green's function, we now analyze the previously eluded to $z$-dependency of the Wannier functions, and thus also the $z$-dependency of the LDOS without any impurities $\rho^0({\bf{r}}_\parallel,z,\omega)$.
In particular, Fig.~\ref{FigDOS0_sandd_Wannier} presents the LDOS evaluated at ${\bf{r}}_\parallel=0$, i.e., directly above a lattice site, for several tip-sample distances $z$ and for different superconducting pairing symmetries. The selected distances are chosen to approximate experimentally relevant STM tip positions. Since the Wannier functions decay exponentially with increasing $z$, the LDOS exhibits a corresponding overall reduction in amplitude. Consequently, to ensure meaningful comparison between different heights, the LDOS is normalized to the left shoulder of the higher energy coherence peak. Figs. (a)+(b) are normalized to the $\beta$-band peak using the model where the gap on the $\beta$-band is larger than the $\alpha$-band value, while (c)+(d) use the model where the roles are reversed leading to a normalization to the $\alpha$-band peak. 

Most importantly, a clear trend emerges independent of the $s_\pm$- and $d$-wave pairing symmetries. For (a)+(b) the spectral weight of the $\alpha$-band coherence peak at $7\,$meV decreases significantly faster than that of the $\beta$-band coherence peak at $19\,$meV with increasing distance $z$. Although both bands are hybridized and contain contributions from both orbitals, the $\alpha$-band LDOS is more strongly suppressed with increasing height. At larger distances, this results in a substantially reduced $\alpha$-band coherence peak, which may provide a natural explanation for the weak shoulder observed in STM experiments \cite{Hai-Hu-Wen-2025-STMSpm}.
For the reversed role situation (c)+(d) the exact opposite behavior is observed. The coherence peak at $19\,$meV of the $\alpha$-band decreases in weight with increasing distance compared to the $\beta$-band value at $7\,$meV.
This means in the experiment one can discern which coherence peak belongs to which band by varying the distance and observing the comparative effects on the coherence peak for the incipient $\gamma$-band case.

A comparison with the corresponding lattice results shown in Fig.~\ref{DOSLatticeInciFig1} reveals notable differences. At short distances, the Wannier-based calculation yields a considerably larger $\alpha$-band weight, which gradually evolves toward a spectral shape comparable to the lattice result at approximately $z \approx 3\,\text{\AA}$. We emphasize that incorporating an orbital filter directly at the lattice level does not quantitatively reproduce the Wannier-based findings. While the qualitative trends remain consistent, suppression effects are less pronounced in the lattice description, particularly at small distances where the $3d_{x^2-y^2}$-orbital still carries substantial weight. Consequently, the lattice LDOS does not exhibit a strong enhancement of the $\alpha$-band peak relative to the $\beta$-band, as observed in the Wannier LDOS. This distinct behavior originates from the specific spatial structure of the Wannier orbitals shown in Fig.~\ref{Wannier3dAndCutsFig}(c,d).
The Fourier transform of the $3d_{x^2-y^2}$-orbital exhibits vanishing spectral weight along the diagonal directions defined by $k_x = k_y$, while displaying pronounced weight along the high-symmetry lines $k_i = 0$. Notably, the $\alpha$-band is predominantly located along these vertical and horizontal directions in momentum space, whereas the $\beta$-band has lower presence there. This momentum-selective structure of the orbital therefore provides a natural explanation for the behavior observed in the LDOS curves.

Consequently, the $z$-dependent analysis of the density of states constitutes a powerful tool to distinguish between the $\alpha$- and $\beta$-bands and to resolve their respective orbital character in the experiment. Furthermore, we notice that $d$-wave  and $s$-wave symmetries appear almost indistinguishable within the continuum Green's function local density of states, which is due to very orbital selective matrix elements, that suppress the nodal quasiparticle contribution.

\subsection{Crossing $\gamma$-band}
\label{sec:Crossing_Gamma}
\begin{figure}
	\includegraphics[width=0.99\linewidth]{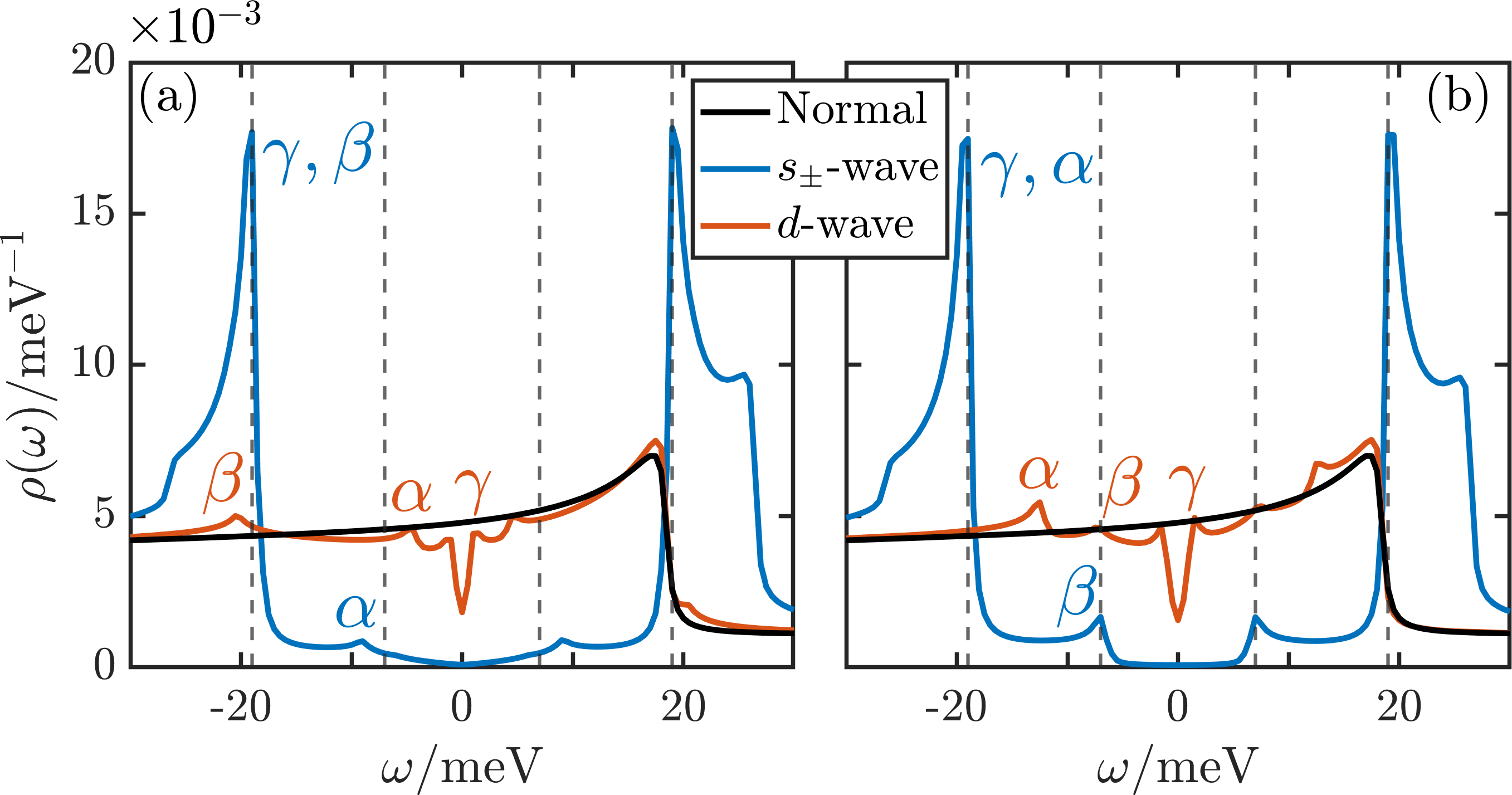}
	\caption{(a) DOS calculated on the lattice in absence of any impurities for the crossing $\gamma$-band case and for the two candidate gap symmetries portrayed in Fig.~\ref{FigGapOnBandsDiffSym}, where the $\beta$-band produces a coherence peak for larger energies than the $\alpha$-band. (b) Vice versa situation where the $\alpha$-band peak is larger than the $\beta$-band peak. The dashed lines are guides to the $7$ and $19\,$meV values reported in~\cite{Hai-Hu-Wen-2025-STMSpm}.}
\label{DOSLatticeCrossFig}
\end{figure}

For the crossing $\gamma$-band scenario, Fig.~\ref{DOSLatticeCrossFig} again presents the lattice-level calculation. In comparison to the incipient case, the resulting curves exhibit notable differences, particularly for the $d$-wave pairing symmetry. For both pairing symmetries, the coherence peak positions differ slightly from the initially proposed values of 19 and 7\,meV. This shift originates from the introduction of the small chemical potential, which modifies the band-projected gap values for the same underlying pairing model. Additionally, the gap on the $\gamma$-band is very small in the $d$-wave case, matching recent theoretical predictions~\cite{Zhan-Hu-2025-Bilayer-FRG-NonlocalCoulomb-SDW-CDW,Maier-Dagotto-2025-Bilayer-InterlayerPairing-Sus}. Consequently, the $d$-wave DOS remains relatively flat and largely follows the normal state DOS, which is dominated by the Van Hove singularity of the $\gamma$-band located at approximately $18\,$meV. The residual spectral weight at $\omega = 0$ in the $d$-wave DOS arises through numerical broadening, which has a magnitude comparable to the small gap on the $\gamma$-band. Already at the lattice level, we see that from the calculated crossing $\gamma$-band spectra only the $s_\pm$-wave scenario yields qualitative agreement with the experimental observations. This result remains unchanged even if the $\gamma$-band coherence peak is shifted to higher energies for the $d$-wave symmetry, as the resulting DOS then presents the characteristic V-shaped profile of a nodal superconductor. Within the $s_\pm$-wave scenario, the construction of the pairing models causes the coherence peak associated with the $\gamma$-band to overlap with either the $\beta$- or $\alpha$-band coherence peak. Both configurations, in which either the $\alpha$- or $\beta$-band produces the dominant coherence peak together with the $\gamma$-band contribution, show comparable agreement with the experiment. Nevertheless, the configuration in which the $\beta$-band hosts the larger gap value appears to provide a more accurate description of the experimental data, since the low energy coherence peak associated with the $\alpha$-band is less pronounced than the corresponding $\beta$-band peak in the alternative configuration.

\begin{figure}
	\includegraphics[width=1\linewidth]{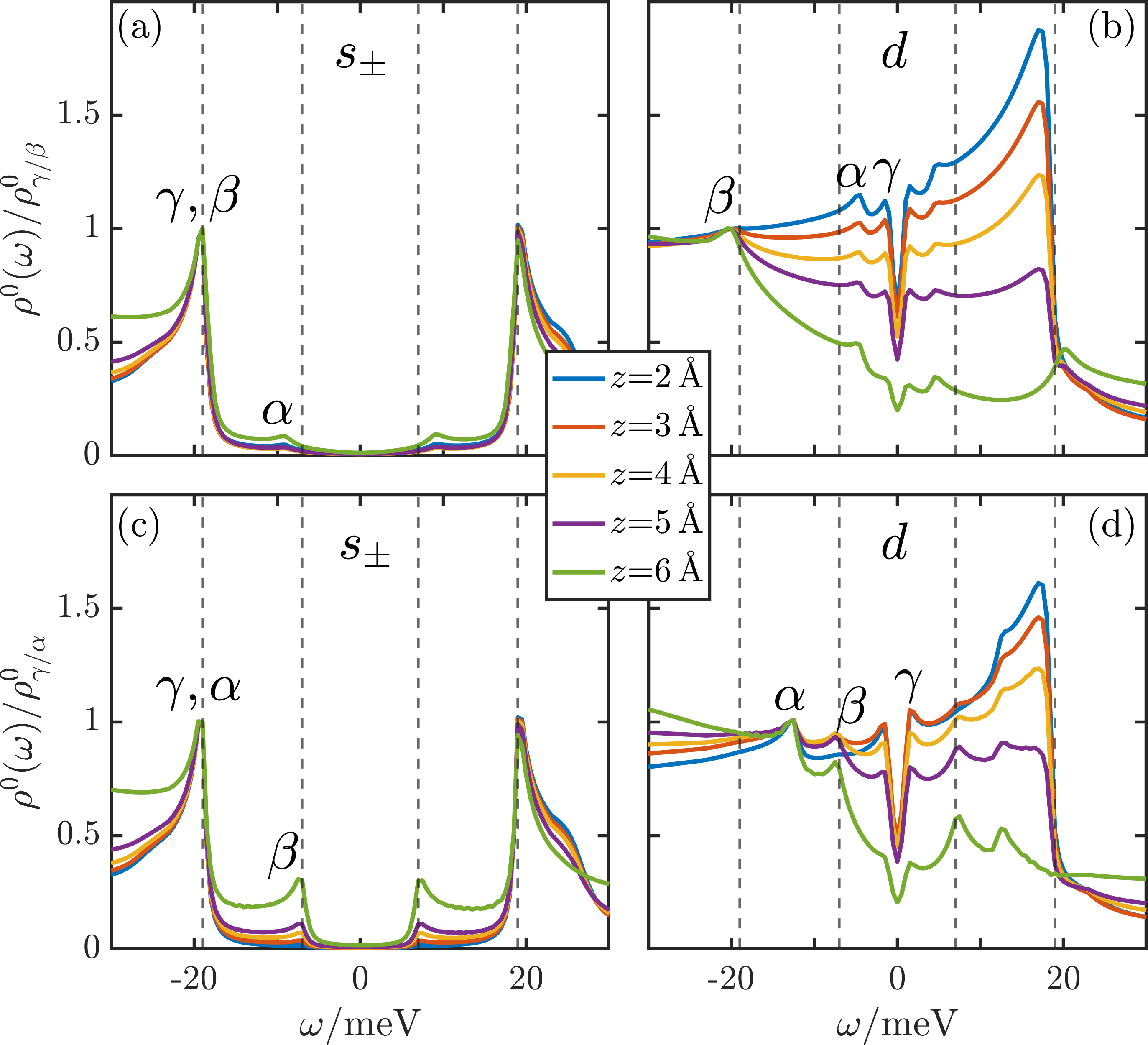}
	\caption{(a)+(b) LDOS normalized by the left shoulder $\beta$-band coherence peak calculated for the crossing $\gamma$-band cases shown in Fig.~\ref{FigGapOnBandsDiffSym} with larger $\beta$-band gap compared to the $\alpha$-band, and without any impurities in dependency on the distance from the top layer of the sample located at $z=0$. The STM tip is directly placed over the lattice site. (c)+(d) Show the reversed situation of a smaller $\beta$-band gap compared to the $\alpha$-band. (a)+(c) Show the evolution of the LDOS for the $s_\pm$-wave symmetry while (b)+(d) show the $d$-wave symmetry.}
\label{FigDOS0_sandd_Wannier_crossing}
\end{figure}

Considering now the LDOS and its dependence on the distance $z$ from the top layer, Fig.~\ref{FigDOS0_sandd_Wannier_crossing} shows that, similar to the incipient case, the $d$-wave scenario still fails to reproduce the experimental observations. For the $s_\pm$-wave symmetry, the dependence on the distance $z$ is considerably weaker than in the incipient $\gamma$-band case. Only in the configuration where the $\beta$-band hosts the smaller gap does the corresponding coherence peak gain considerable spectral weight with increasing distance.
This can be attributed to the dominant contribution of the $\gamma$-band, which is primarily composed of $3d_{z^2}$ orbital character. Therefor, variations in distance lead only to comparatively small changes in the LDOS. Overall, the crossing $\gamma$-band scenario exhibits only a weak dependence on the tip-sample distance, while still reproducing the experimentally observed low energy coherence peak, regardless of the band origin.

\section{Quasiparticle interference}
\label{sec:QPI}
Including the effect impurities in our consideration, we follow the $\hat{T}$-matrix formalism to calculate the correction to the local density of states due to impurity scattering of a single impurity site on the lattice level
\begin{equation}
	\begin{split}
	   \delta\rho({\bf{q}},\omega)=-\frac{2}{\pi}\sum_{\bf{k}}\Im\Tr\frac{\tau_0+\tau_z}{2}\hat{L}_1\hat{G}_{\bf{k}}(\omega)\hat{T}(\omega)\hat{G}_{\bf{k}+\bf{q}}(\omega),
	\end{split}
	\label{QPILatticeFormula}
\end{equation}
The essential quantity defined above is the $\hat{T}$-matrix given by $\hat{T}(\omega)=[\mathbb{1}-\hat{V}\hat{G}(\omega)]^{-1}\hat{V}$ where $\hat{G}(\omega)=\sum_{\bf{k}}\hat{G}_{\bf{k}}(\omega)$ is the momentum integrated Green's function and $\hat{V}$ is the impurity matrix. For the impurity matrix $\hat{V}$ we now have multiple choices, depending on where the impurity is placed in our sample. 

The obvious choices are a placement on either of the two layer's nickel sites further labeled as $\hat{V}_{1/2}=V_0 \tau_z (\lambda_0\pm\lambda_z)\sigma_0/2$ for the respective layer for a non-magnetic impurity, where $V_0$ is the impurity strength and $\lambda_i$ ($\sigma_i$) are the Pauli matrices in layer (orbital) space. As either of the two choices only places the impurity on one of the two layers, the impurity potential breaks the mirror-symmetry of the system. This contrasts Ref.~\cite{Zhang-2025-MirrorSelectiveQPIBilayer}, where the impurity potentials are constructed as linear combinations of the potentials shown here, yielding overall potentials that are either even or odd under mirror-symmetry. However, since we are still subject to the layer projection $\hat{L}_1$, the overall mirror-symmetry of the bilayer system remains broken. This mirror-symmetry breaking leads to the scattering channels between bonding-bonding (bb), antibonding-antibonding (aa) and bonding-antibonding (ba) states to mix in the overall LDOS measured by the experiment. 
While it is easy to separate the channels in theoretical calculations, the experiment can only do so under ideal conditions. Under the assumption of identical impurities with the same impurity strength situated in different layers with respective impurity potentials $\hat{V}_{1/2}$, one can measure them individually and afterwards build linear combinations of the measured LDOS. For $\hat{V}_1+\hat{V}_2$ the remnant signal corresponds to the mirror-even bb and aa channels, while for $\hat{V}_1-\hat{V}_2$ one is left with the mirror-odd ba channel.

Another choice for the impurity matrix is to model an impurity at the apical oxygen site between the two layers, henceforth labeled $\hat{V}_3=V_0 \tau_z \lambda_x (\sigma_0-\sigma_z)/2$. We see that the chosen structure is inherently different from $\hat{V}_{1/2}$, as it includes not only interlayer coupling $\lambda_x$, but is also orbital-selective. These two features are the result of the characteristic shapes of the $3d_{x^2-y^2}$ and $3d_{z^2}$ orbitals, where the former is drastically more planar than the latter. This in turn leads to a stronger interlayer scattering of the $3d_{z^2}$ than the $3d_{x^2-y^2}$-orbital. Note that the here defined impurity potential $\hat{V}_3$ is mirror-symmetric, because the impurity is situated exactly in the middle of the two layers. Following the previous discussion of linear combinations of $\hat{V}_1$ and $\hat{V}_2$, this mirror symmetry then automatically leads to a decoupling of the QPI signal into only the bb and aa channels, giving no contribution in the ba channel.

It further proves useful to define the antisymmetrized version of the correction to the LDOS from Eq. (\ref{QPILatticeFormula}), measuring the particle-hole asymmetry of our system. Following the HAEM recipe \cite{HAEM-2015-Original,HAEM-2018-QPILiFeAs} we define it as
\begin{equation}
	\begin{split}
	   \delta\rho^-({\bf{q}},\omega)=\delta\rho({\bf{q}},\omega)-\delta\rho({\bf{q}},-\omega).
	\end{split}
	\label{QPIHAEMLatticeFormula}
\end{equation}

\begin{figure}
	\includegraphics[width=1\linewidth]{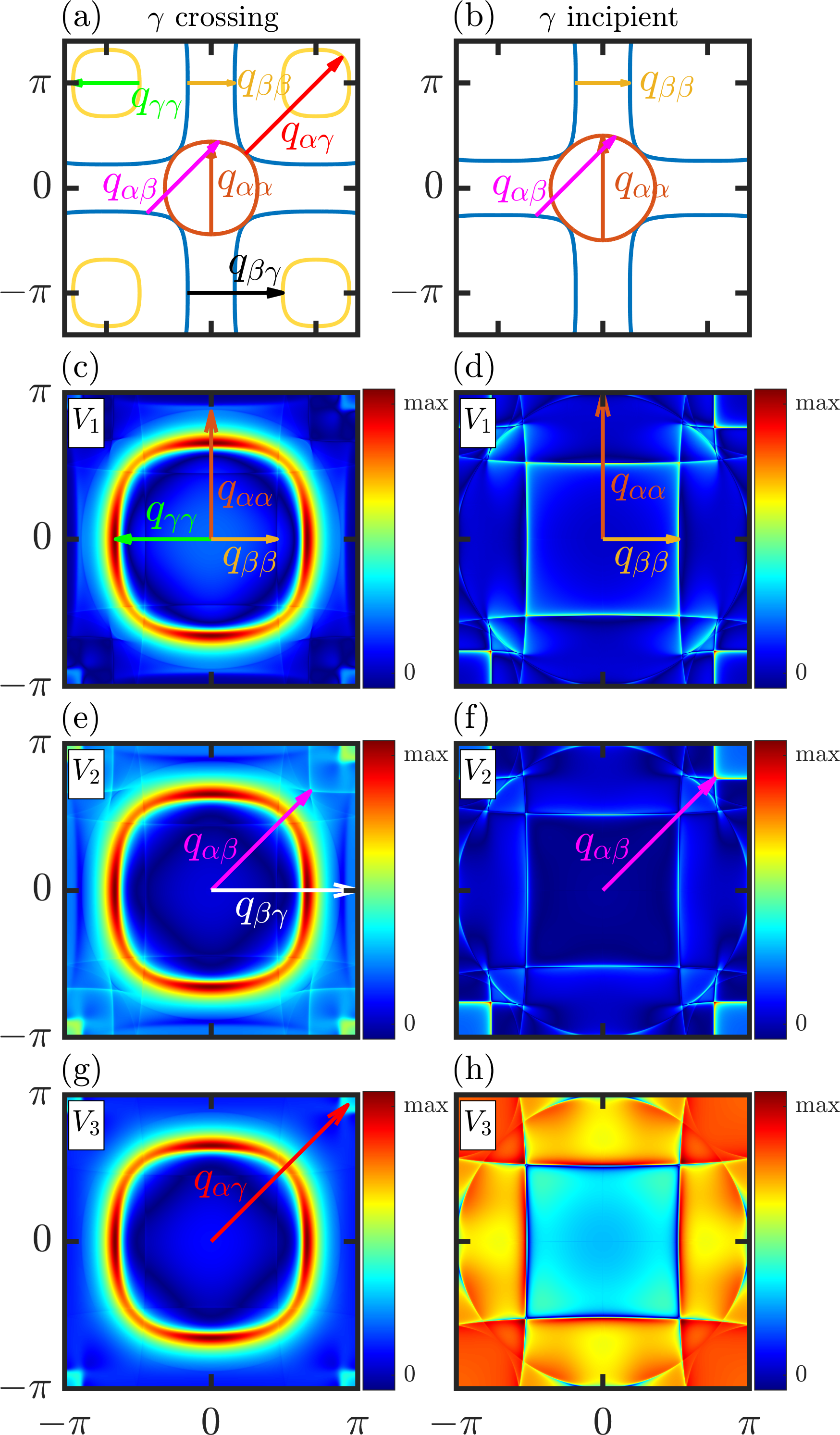}
	\caption{Correction to the density of states $\abs{\delta\rho({\bf{q}},\omega=0)}$ in the normal state. The rows are showing the different impurity potentials $\hat{V}_i$ and the columns discern the case of a crossing and incipient $\gamma$-band. The QPI pattern is calculated for a weak non-magnetic impurity ($V_0=-10$ meV). Color scale units are normalized to the maximum value of the corresponding plot.}
\label{QPINormalFig}
\end{figure}

Before we go into the details of the superconducting system, let us first analyze the features of the normal state. Using Eq. (\ref{QPILatticeFormula}) we calculate the QPI patterns at $\omega=0$ for a weak non-magnetic impurity described by the different impurity potentials $\hat{V}_i$ with $V_0=-10\,$meV. Fig.~\ref{QPINormalFig} distinguishes between the crossing and the incipient $\gamma$-band cases by comparison in the columns. The figure shows the total QPI patterns $\abs{\delta\rho({\bf{q}},\omega=0)}$ of Eq. (\ref{QPILatticeFormula}), which means that they are not separated into channels. For the bb channel we find the scattering vectors $\bf{q}_{\alpha\alpha}$, $\bf{q}_{\gamma\gamma}$ and $\bf{q}_{\alpha\gamma}$, where the latter are not present in the incipient case because of the absence of the $\gamma$-band at the Fermi level. This absence also triggers the disappearance of the $\bf{q}_{\beta\gamma}$ vector in the ba channel. Otherwise, the ba channel only hosts a scattering $\bf{q}_{\alpha\beta}$. Similarly, the aa channel hosts one scattering $\bf{q}_{\beta\beta}$ since the $\beta$-band is the only antibonding band that crosses the Fermi level.
Generally comparing the incipient with the crossing case, we see that the main feature associated with the crossing $\gamma$-band is the ring-shaped scattering of the $\gamma$-band with itself.
This means QPI is an excellent experimental tool to probe the incipiency of the $\gamma$-band in bilayer nickelate thin films.
As for the comparison of the impurity potentials, as previously mentioned a linear combination of the $\hat{V}_{1/2}$ leads to a partial decomposition into even and odd symmetry channels. In turn this means that their individual features are not so different from each other, making an experimental distinction hard, which contrasts the clear decomposition into symmetry channels made in Ref.~\cite{Zhang-2025-MirrorSelectiveQPIBilayer}. 
Looking at the apical oxygen impurity $\hat{V}_3$, one sees that it displays the same features with differing intensities. Its orbital selectivity leads to the $\gamma$-band features, which are almost exclusively of $d_{z^2}$-character, to shadow the other contributions in the crossing case. The incipient case shows a redistribution of weight to the outside of the rectangle described by the aa scattering of the $\beta$-band. In the experiment this could be a good feature to probe for the incipiency of the $\gamma$-band and presence of an apical oxygen impurity, because all other cases do not display this asymmetry and instead show high intensity lines.

Turning to the QPI signals in the superconducting state, all arguments concerning channel splitting that were established for the normal state remain valid in the presence of superconductivity. In contrast to the normal state analysis, where we focused on the spectra of $\abs{\delta\rho({\bf q},\omega=0)}$, we now consider finite energies $\omega$ in order to identify scattering processes between the superconducting gap openings on the Fermi surface.
In this context, it is advantageous to analyze the antisymmetrized QPI signal $\delta\rho^-({\bf q},\omega)$, introduced in Eq. (\ref{QPIHAEMLatticeFormula}), which provides a direct measure of the particle-hole asymmetry of the underlying superconducting state.

\begin{figure*}[t]
    \centering
    \includegraphics[width=0.9\textwidth]{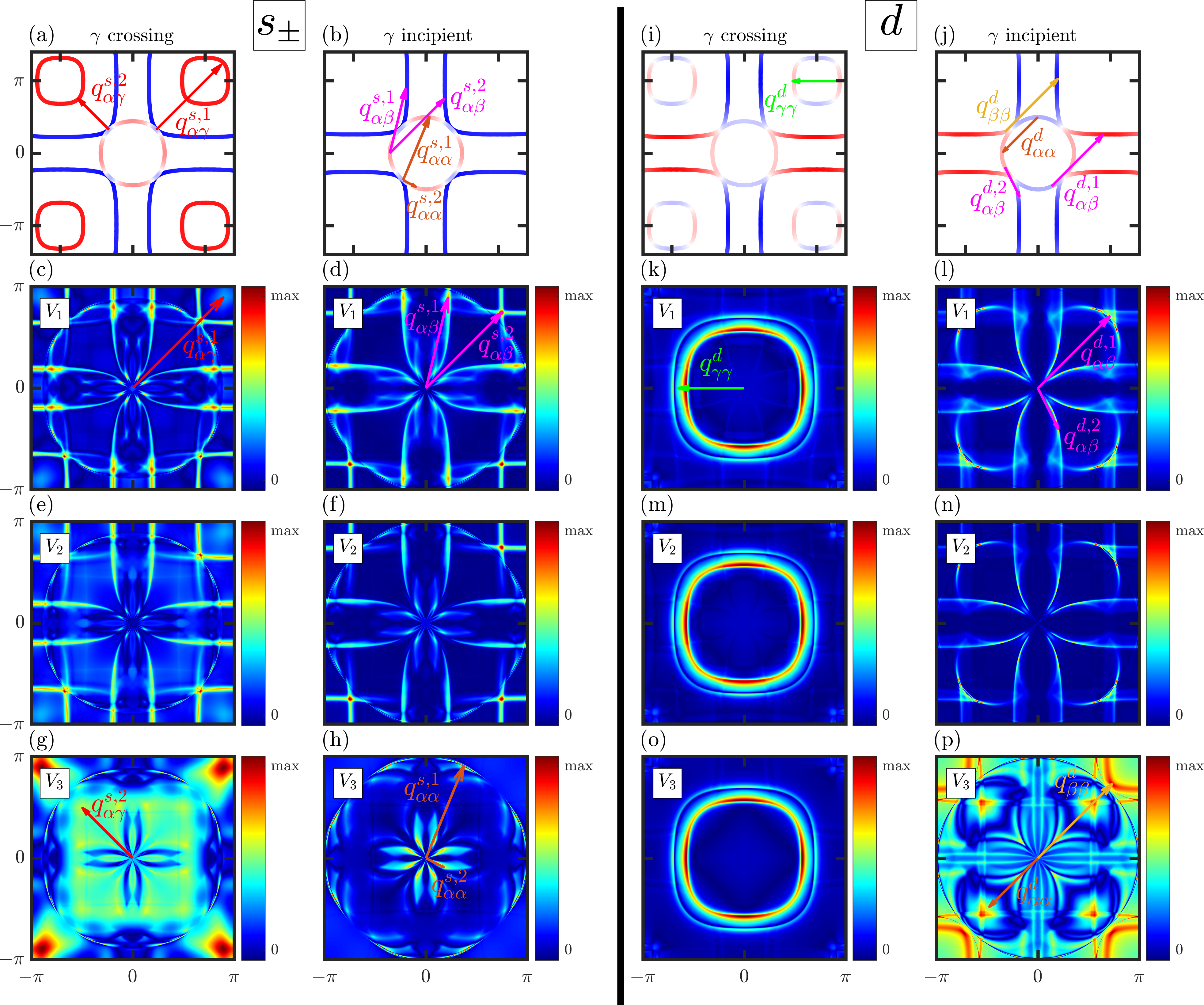}
    \caption{Antisymmetrized correction to the density of states $\abs{\delta\rho^-({\bf{q}},\omega=5\,\textrm{meV})}$ in the superconducting state for the $s_\pm$- and $d$-wave symmetry and for the incipient and crossing $\gamma$-band scenarios. The rows discern between the different impurity potentials $\hat{V}_i$, where the first row is reserved for the projected gap on the Fermi surface including the scattering vectors connecting different scattering events that are present in the QPI patterns. The QPI pattern is calculated for a weak non-magnetic impurity ($V_0=-10$ meV). Color scale units are normalized to the maximum value of the corresponding plot. }
    \label{FigQPISCLatticeAll}
\end{figure*}

Fig.~\ref{FigQPISCLatticeAll} shows QPI signatures for the two candidate gap symmetries in the incipient and crossing $\gamma$-band case, considering impurity potentials $\hat{V}_{1/2}$ and $\hat{V}_3$. As already observed in the normal state analysis, it is sufficient to focus on impurity potential $\hat{V}_1$, since $\hat{V}_2$ yields qualitatively identical QPI patterns, differing only in their relative intensities.  In contrast to Ref.~\cite{Zhang-2025-MirrorSelectiveQPIBilayer}, our pairing model was chosen such that it suits the experimentally measured DOS, and thus gives a more rich and realistic view in terms of expected weights for the QPI signal, leading to additional features and shifted weights.

In the crossing $\gamma$-band case, the presence of the $\gamma$-band gives rise to QPI features that closely resemble those of the normal state for the $d$-wave gap symmetry. This behavior can be attributed to the very small superconducting gap on the $\gamma$-band in the chosen model. For the $s_\pm$-wave symmetry, contributions associated with the $\gamma$-band are generally speaking small for energies below the pairing strength and resemble the normal state peaks outlined in Fig.~\ref{QPINormalFig}, with the exception of ${\bf{q}}_{\alpha\gamma}^{s,2}$.

A closer analysis of Fig.~\ref{FigQPISCLatticeAll} with exclusion of the crossing $\gamma$ band for the $d$-wave pairing state, we find that for impurity potentials $\hat{V}_{1/2}$ the dominant QPI features originate from $\alpha$-$\beta$ scattering and thus from the ba channel for both gap symmetries. Contributions from the bb and aa channels are comparatively weak and partially overlap with signals arising from the ba channel. As a consequence, it becomes experimentally challenging to distinguish between the two gap symmetries, since both exhibit features at similar wave vectors and with comparable structure.
The only feature without a clear counterpart between the two gap symmetries is the star-shaped fringe pattern associated with the scattering vector ${\bf{q}}_{\alpha\alpha}^{s,2}$. This feature appears exclusively in the $s_\pm$-wave case and arises from the nodal structure of the $\alpha$-band in our model, which hosts accidental nodes. Although this signal is most pronounced for the impurity potential $\hat{V}_3$, which does not host any ba contributions, it is also slightly visible for $\hat{V}_{1/2}$. This observation suggests that QPI provides a sensitive probe of the nodal structure in the system's strongly hybridized bands.
For the $d$-wave symmetry and an incipient $\gamma$-band in the presence of an apical oxygen impurity, the absence of the ba channel causes Fig.~\ref{FigQPISCLatticeAll}(p) to display $\beta$-$\beta$ scattering contributions from the aa channel to dominate the QPI response. A comparison between the $d$-wave and $s_\pm$-wave cases for this impurity reveals clear qualitative differences that should be experimentally accessible, in contrast to their largely similar signatures for impurity potential $\hat{V}_{1/2}$. We note that the feature associated to the scattering vector ${\bf{q}}_{\alpha\alpha}^{d}$ is sub-leading and does not correspond to the rectangular pattern of the $\beta$-$\beta$ scattering, but to the circular intensity increase in the background along the diagonals.

\begin{figure}
	\includegraphics[width=0.59\linewidth]{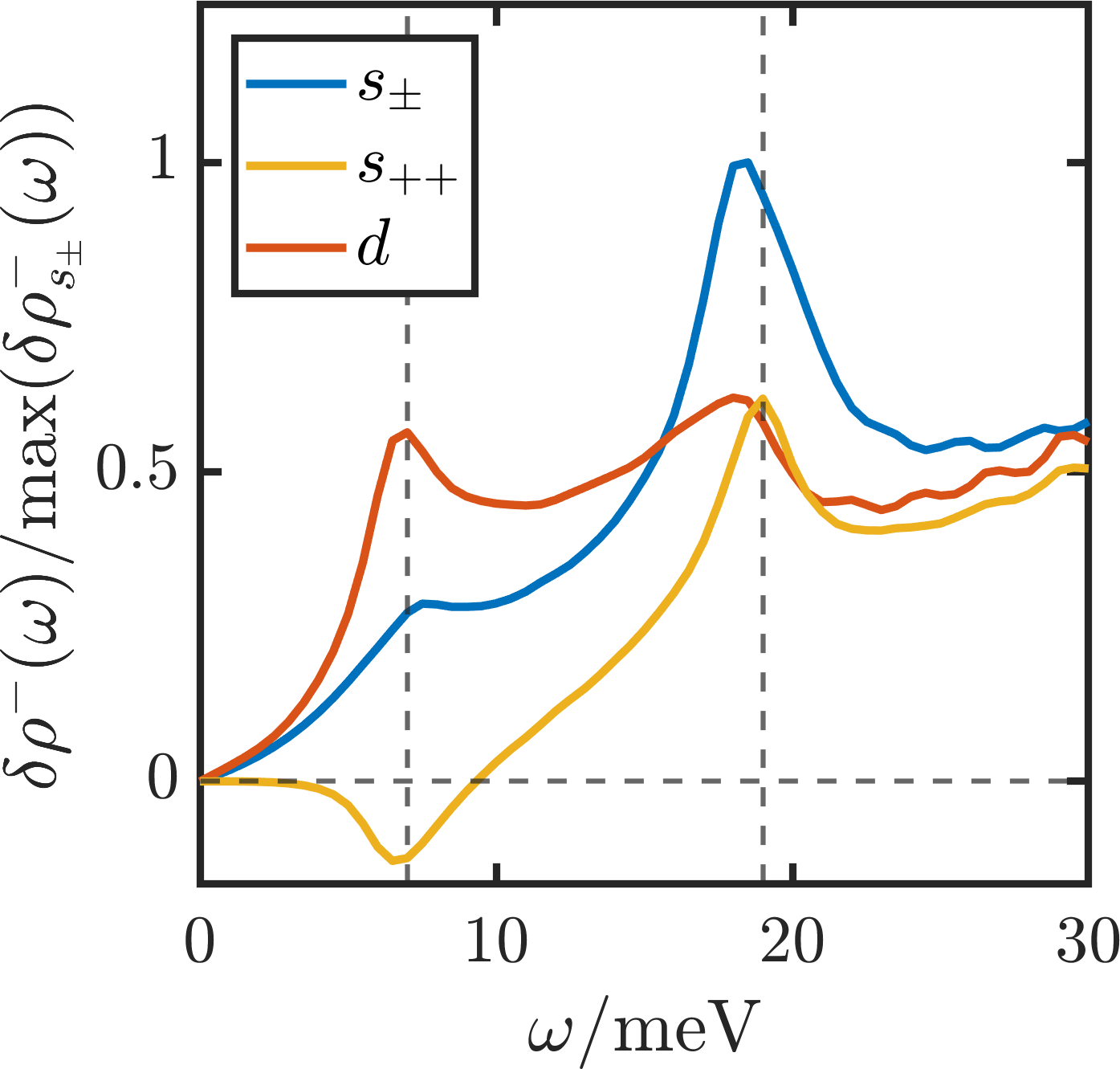}
	\caption{Calculated momentum integrated antisymmetrized correction to the DOS using the HAEM prescription \cite{HAEM-2015-Original,HAEM-2018-QPILiFeAs} with a weak impurity $V_0=-10\,$meV and potential $\hat{V}_1$. The dashed guidelines show the experimentally reported values of $7$ and $19\,$meV \cite{Hai-Hu-Wen-2025-STMSpm}.The curves are normalized with respect to the maximum of the blue curve corresponding to the $s_\pm$-wave symmetry.}
\label{HAEMLatticeInciFig}
\end{figure}

In the next step we are now looking into the HAEM method \cite{HAEM-2015-Original,HAEM-2018-QPILiFeAs} calculating the momentum integrated correction to the DOS $\delta\rho^-(\omega)$, a quantity sensitive to the phase of the superconducting order parameter. Fig.~\ref{HAEMLatticeInciFig} (b) compares the signals for our proposed $s_\pm$ and $s_{++}$ gaps with the expected result of a sign change for the $s_{++}$ symmetry and a uniform sign for the $s_\pm$ symmetry. As is typical, the signal of the $s_\pm$ gap is larger than that of the $s_{++}$ gap. A comparison of the $s_\pm$- and $d$-wave symmetries does not yield any distinctive features to probe in the experiment, as both gaps are sign changing and produce a similar curve.
As for the crossing $\gamma$-band scenario, the plots look very similar, just with a very dominant peak at $19\,$meV because of the high weight of the $\gamma$-band. In total, this means the HAEM method is an ideal tool to probe the phase of the superconducting gap and confirm the predicted sign changing behavior of the $s_\pm$ or $d$-wave symmetry.

\begin{figure}
	\includegraphics[width=1\linewidth]{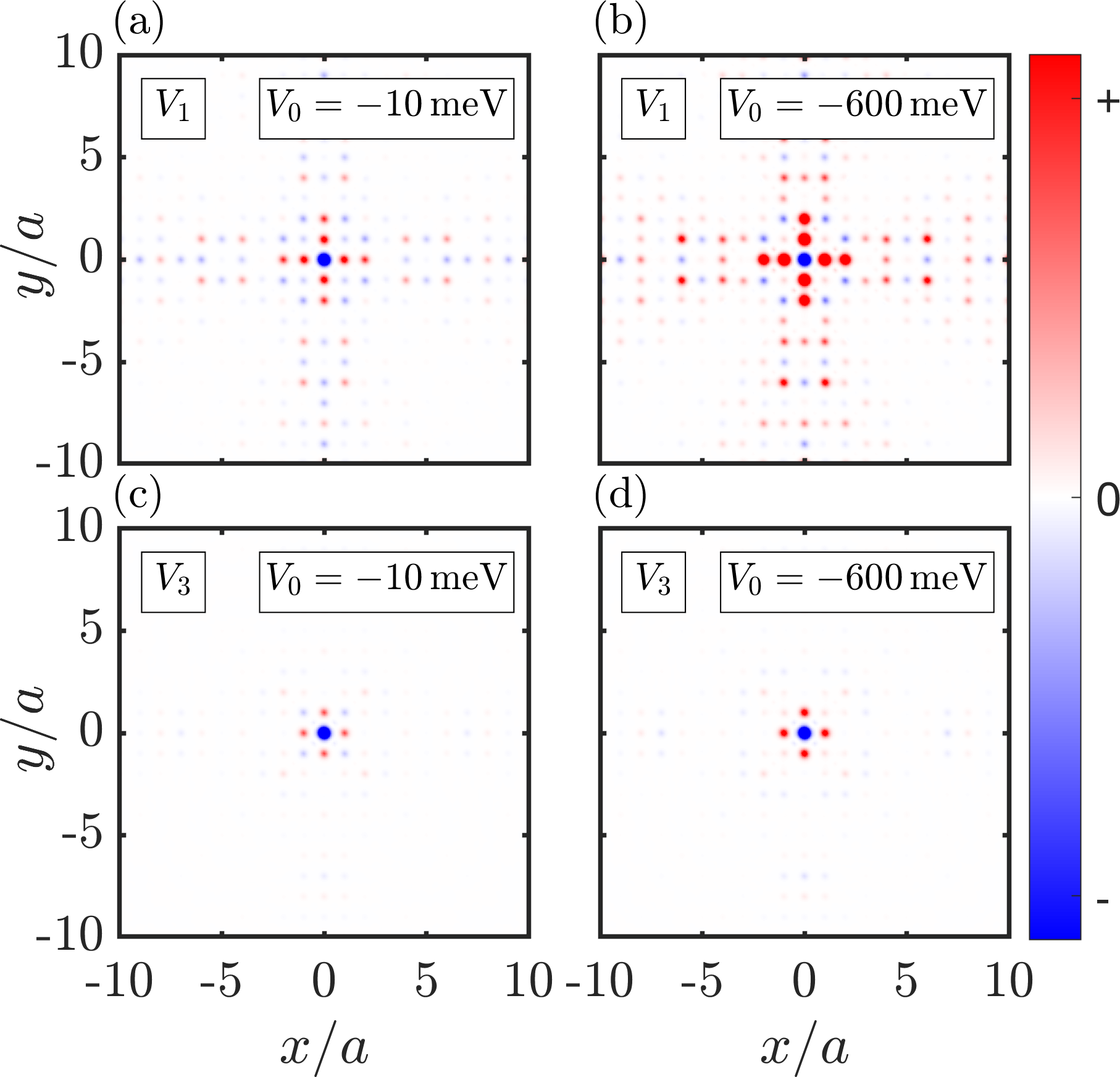}
	\caption{Real space correction to the LDOS $\delta\rho({\bf{r}},\omega=5\,\textrm{meV})$ for the $s_\pm$-wave symmetry and an incipient $\gamma$-band at a STM tip distance $z\approx3\,$\AA. Columns discern between a weak ($V_0=-10\,$meV) and strong ($V_0=-600\,$meV) impurity, while rows differentiate between the impurity potentials $\hat{V}_1$ and $\hat{V}_3$.}
\label{FigQPIRealSpace}
\end{figure}

Next, let us analyze the corrections to the LDOS through impurity scattering $\delta\rho(\bf{r},\omega)$ in the continuum using the Wannier weighted QPI spectra. The real space profiles for a weak ($V_0=-10\,$meV) and strong ($V_0=-600\,$meV) impurity in the top layer ($\hat{V}_1$) and at the apical oxygen ($\hat{V}_3$), at a distance $z \approx 3\,\text{\AA}$ and for a $s_\pm$-wave symmetry in the incipient $\gamma$-band case are shown in Fig.~\ref{FigQPIRealSpace}. The resulting pattern has an approximate cross shaped form, where the individual spots are almost atomic-like. This atomic-like shape is rooted in the strong localization of the $3d_{z^2}$-orbital, as shown in Fig. \ref{Wannier3dAndCutsFig} (d). Increasing the impurity strength yields a qualitatively similar picture, except that the cross shape gets larger and covers even more lattice sites. In addition to the cross feature, slight wave fringes appear spreading out of it. A change to the $d$-wave symmetry does not yield any significant changes to the real space profile.

\begin{figure}[t]
	\includegraphics[width=1\linewidth]{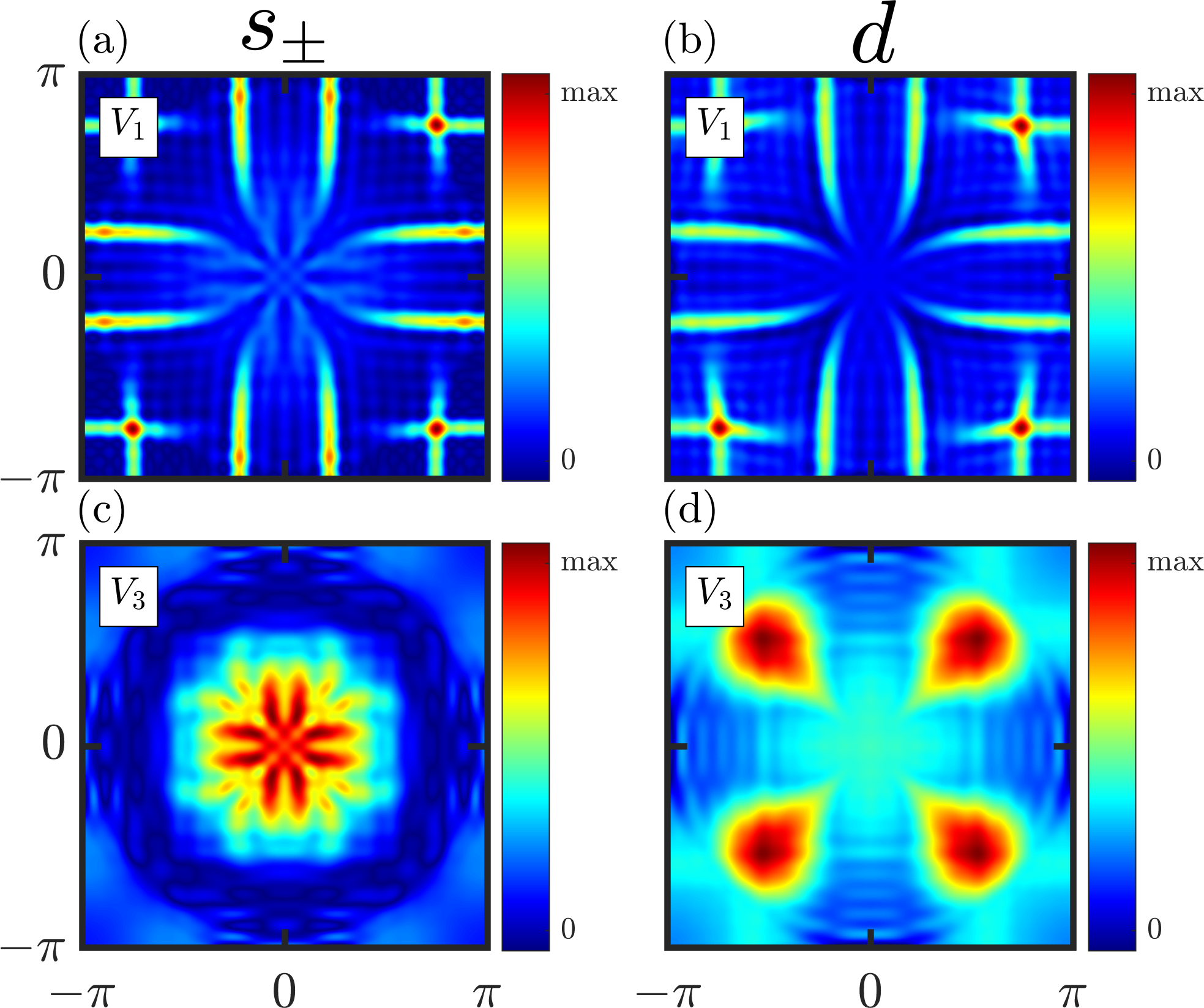}
	\caption{Wannier adjusted antisymmetrized correction to the density of states $\abs{\delta\rho^-({\bf{q}},\omega=5\,\textrm{meV})}$ in the superconducting state for the $s_\pm$- and $d$-wave symmetry and for an incipient $\gamma$-band. Shown is the continuum Green's function analogue to the lattice calculation of Fig.~\ref{FigQPISCLatticeAll}. (a)+(b) depict the impurity potentials $\hat{V}_1$ with an impurity in the top layer and (c)+(d) show $\hat{V}_3$ the apical oxygen impurity potential. The QPI pattern is calculated for a weak non-magnetic impurity ($V_0=-10$ meV) and a tip distance of $z\approx3\,${\AA} from the top layer.}
\label{FigQPISCWannierSelected}
\end{figure}

Fourier transformation of the real-space LDOS corrections into momentum space yields the spectra presented in Fig.~\ref{FigQPISCWannierSelected}, where we again display the antisymmetrized correction $\delta\rho^-({\bf{q}},\omega)$.  A comparison of these continuum results with their lattice counterparts shown in Fig.~\ref{FigQPISCLatticeAll} reveals that the overall features remain qualitatively consistent across both approaches.
For the impurity potential $\hat{V}_1$, the QPI spectra remain largely unchanged relative to the lattice formulation, with dominant contributions arising from interband $\alpha$-$\beta$ scattering processes in the ab channel. More pronounced modifications occur for the apical oxygen impurity $\hat{V}_3$. In this case, the scattering processes in the bb channel associated with the $\alpha$-band become significantly enhanced. In the $s_\pm$ pairing state, this enhancement manifests as the pronounced star-shaped feature corresponding to the scattering vector ${\bf{q}}_{\alpha\alpha}^{s,2}$ of the accidental nodes, which now dominates over ${\bf{q}}_{\alpha\alpha}^{s,1}$.
The modifications are even more substantial for the $d$-wave symmetry. Previously, the $\beta$-$\beta$ scattering constituted the leading contribution, but now the continuum formalism instead emphasizes the circular $\alpha$-$\alpha$ scattering features, which were sub-leading in the lattice calculation. This difference between the $s_\pm$- and $d$-wave scattering patterns in the presence of the apical oxygen impurity highlights the sensitivity of QPI to the underlying gap symmetry, thereby providing a robust tool to differentiate between these two gap structures.

As for an analysis of the STM tip distance influencing the QPI spectra, we find only insignificant changes for the experimental verification. 
Similar to the LDOS without impurities, the spectral weight of the individual features changes. Again, for smaller distances the $\alpha$-$\alpha$ scattering becomes more pronounced in comparison to the other features, making the star-shape of the $s_\pm$-symmetry less dim for the impurity potential $\hat{V}_1$ and increasing the weight of the circular features of the $d$-wave for the impurity potential $\hat{V}_3$ even more.

\section{Summary}
\label{sec:Summary}
In summary, motivated by the recent advances of the STM experiments \cite{Hai-Hu-Wen-2025-STMSpm,liang-2026-STM-LaPr-Thin-UShape,wang-2026-STM-LaPr-Thin-UShape2Peaks}, we analyzed the superconducting STM spectra in bilayer nickelates by using continuum Green's functions, which allow us to investigate the spatial dependence of the local density of states as a function of STM tip height. We showed that the distinct spatial profiles of the $3d_{x^2-y^2}$ and $3d_{z^2}$ orbitals lead to a pronounced orbital filtering effect, strongly modifying the relative visibility of $\alpha$- and $\beta$-band coherence peaks. This $z$-dependent analysis naturally explains the experimentally weak $\alpha$-band shoulder and demonstrates that continuum modeling is essential for quantitative comparison with STM data. Furthermore, we propose that experimental studies with varying tip height may add important information on the symmetry of the superconducting order parameter in bilayer nickelates. 

Moreover, Wannier-resolved real- and momentum-space quaiparticle interference patterns exhibit atomically localized features rooted in the orbital character, further emphasizing the importance of orbital-selective effects, improving upon prior results of Ref.~\cite{Zhang-2025-MirrorSelectiveQPIBilayer}. For impurities located on a single layer, bonding-antibonding channel mixing leads to similar dominant $\alpha$-$\beta$ scattering features for both $s_\pm$ and $d$-wave states, making experimental discrimination challenging. In contrast, mirror-symmetric apical oxygen impurities selectively suppress bonding-antibonding contributions and enhance channel-resolved features, exposing qualitative differences between $s_\pm$ and $d$-wave pairing. In particular, star-shaped scattering features associated with accidental $\alpha$-band nodes emerge uniquely in the $s_\pm$ case, highlighting the sensitivity of QPI to subtle nodal structures in hybridized multiorbital systems. 
Using the antisymmetrized HAEM prescription, we demonstrated that the momentum-integrated QPI response robustly detects the sign change of the superconducting order parameter. The clear contrast between sign-changing $s_\pm$- and sign-preserving $s_{++}$-gaps confirms the suitability of this method for experimentally establishing the phase structure of the pairing state.

Overall, our results establish a unified framework linking impurity symmetry, superconducting gap structure, and STM observables in bilayer nickelates. We show that a combination of  Wannier-resolved continuum modeling, channel-selective QPI analysis, HAEM antisymmetrization, can provide experimentally accessible criteria to determine the incipient nature of the $\gamma$-band and distinguish between competing superconducting gap symmetries. These findings offer concrete guidance for future STM investigations and contribute to clarifying the pairing mechanism in bilayer nickelate superconductors.


\section{Acknowledgments} 
\label{sec:Acknowledgments}
The work is supported by the German Research Foundation Project Number 572794210.

\appendix
\section{Superconducting gaps}
 \label{appendix_SuperconductingGaps}
 \renewcommand\thefigure{\thesection.\arabic{figure}}  
 \setcounter{figure}{0} 
\begin{table}[h!]
\centering
\begin{tabular}{c||c|c|c}
    & $s_\pm$-wave & $d$-wave& $s_{++}$-wave\\
\hline\hline
$\Delta^{\parallel}_{x,0}$ & $-13.5$ & $-31.5$ & $13.0$\\
\hline
$\Delta^{\parallel}_{z,0}$ & $2.0$ & $-5.4$ & $12.0$\\
\hline
$\Delta^{\perp}_{x,0}$ & $7.5$ & $-0.9$ & $-6.0$\\
\hline
$\Delta^{\perp}_{z,0}$ & $17.0$ & $9.9$ & $-7.0$\\
\end{tabular}
\caption{Intraorbital pairing strength coefficients for the different gap symmetries and orbitals. All values are given in meV.}
\label{TableGapCoeff}
\end{table}

Similar to the previous Table \ref{TableGapCoeff}, we now also construct a table for the pairing model where the roles of the $\alpha$- and $\beta$-band are reversed, yielding the projected $\alpha$-band value of $19\,$meV and $7\,$meV for the $\beta$-band in the incipient $\gamma$-band case. The values for this role reversed model are given in Table \ref{TableGapCoeffSpecialApp}.
\begin{table}[!h]
	\centering
	\begin{tabular}{c||c|c}
		& $s_\pm$-wave & $d$-wave \\
		\hline\hline
		$\Delta^{\parallel}_{x,0}$ & $6.5$ & $-45.3$  \\
		\hline
		$\Delta^{\parallel}_{z,0}$ & $5.5$ & $-2.0$ \\
		\hline
		$\Delta^{\perp}_{x,0}$ & $12.5$ & $-38.8$  \\
		\hline
		$\Delta^{\perp}_{z,0}$ & $13.5$ & $8.0$  \\
	\end{tabular}
	\caption{Intraorbital pairing strength coefficients for the different gap symmetries and orbitals in the special model that reverses the projected $\alpha$- and $\beta$-band gap values. All values are given in meV.}
	\label{TableGapCoeffSpecialApp}
\end{table}

\begin{figure}[h]
    \includegraphics[width=1\linewidth]{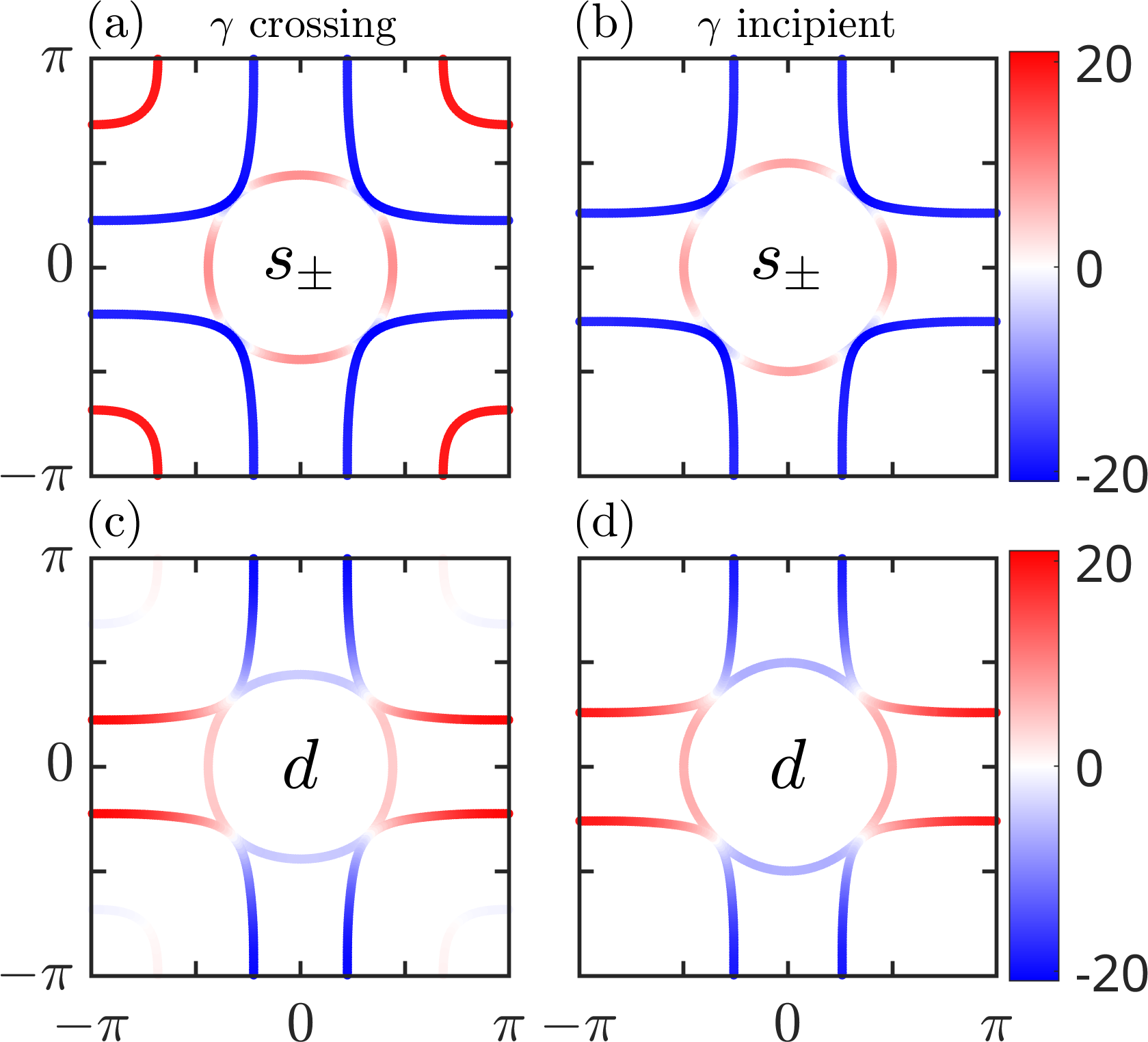}
	\caption{Projected gap on the bands for the different gap symmetries and the crossing and incipient $\gamma$-band case. The units of the colorbars are in meV.}
\label{FigGapOnBandsDiffSym}
\end{figure}

In Fig.~\ref{FigGapOnBandsDiffSym} we show the projected gap onto the Fermi surface for the incipient and crossing $\gamma$-band in the case of the pairing coefficients of Table \ref{TableGapCoeff}.

\bibliography{literature}

\end{document}